\documentclass[letter,twocolumn]{autart}

\usepackage{natbib} 
\usepackage[english]{babel}
\usepackage{amsmath,amssymb,latexsym,amsfonts,mathptmx}
\usepackage{xcolor,graphicx,subfigure}
\usepackage{enumerate}
\usepackage{epsfig}
\catcode`\<=\active \def<{
     \fontencoding{T1}\selectfont\symbol{60}\fontencoding{\encodingdefault}}
\usepackage[normalem]{ulem}
\usepackage{times}

\newtheorem{theorem}{Theorem}[section]

\newtheorem{assumption}[theorem]{Assumption}

\newtheorem{lemma}[theorem]{Lemma}
\newtheorem{corollary}[theorem]{Corollary}

\newtheorem{remark}[theorem]{Remark}
\newtheorem{proposition}[theorem]{Proposition}
\newcommand{\oprocendsymbol}{\hbox{$\bullet$}}
\newcommand{\oprocend}{\relax\ifmmode\else\unskip\hfill\fi\oprocendsymbol}

\newcommand{\nin}{\not\in}

\newcommand{\tmtextit}[1]{{\itshape{#1}}}

\newcommand{\closure}{_{\operatorname{cl}}}
\newcommand{\real}{{\mathbb{R}}}
\newcommand{\naturals}{{\mathbb{N}}}
\newcommand{\range}{\operatorname{range}}

\renewcommand{\theenumi}{(\roman{enumi})}

\newcommand{\zeros}{\bold{0}}
\newcommand{\ones}{\bold{1}}

\newcommand{\diag}{\operatorname{diag}}

\newcommand{\setdef}[2]{\left\{ #1 \; \big| \; #2\right\}}

\newcommand{\CC}{\mathfrak{C}}

\newcommand{\longthmtitle}[1]{\tmtextit{(#1).}}

\allowdisplaybreaks

\begin{document}

\begin{frontmatter}



  \title{Distributed Transient Frequency Control for Power Networks
    with Stability and Performance Guarantees}

  \thanks{A preliminary version has been accepted at the IEEE
    Conference on Decision and Control as~\citep{YZ-JC:18-cdc1}.  This
    work was supported by NSF award CNS-1329619 and AFOSR Award
    FA9550-15-1-0108.}

  \author[First]{Yifu Zhang}%
  \author[First]{\quad Jorge Cort\'es}
  
  \address[First]{Department of Mechanical and Aerospace Engineering,
    University of California, San Diego,
    \{yifuzhang,cortes\}@ucsd.edu}
  
  \begin{abstract}
    This paper proposes a distributed strategy regulated on a subset
    of individual buses in a power network described by the swing
    equations to achieve transient frequency control while preserving
    asymptotic stability. Transient frequency control refers to the
    ability to maintain the transient frequency of each bus of
    interest in a given safe region, provided it is initially in it,
    and ii) if it is initially not, then drive the frequency to
    converge to this region within a finite time, with a guaranteed
    convergence rate.  Building on Lyapunov stability and set
    invariance theory, we formulate the stability and the transient
    frequency requirements as two separate constraints for the control
    input. Our design synthesizes a controller that satisfies both
    constraints simultaneously.  The controller is distributed and
    Lipschitz, guaranteeing the existence and uniqueness of the
    trajectories of the closed-loop system.  We further bound its
    magnitude and demonstrate its robustness against measurement
    inaccuracies.  Simulations on the IEEE 39-bus power network
    illustrate our results.
  \end{abstract}

  
\end{frontmatter}

\section{Introduction}

In power system planning and operation against contingencies (e.g.,
generator loss, transmission line tripping, unexpected power demands),
to avoid the system from running underfrequency or to help the network
recover from it, load shedding and curtailment are commonly employed
to balance supply and demand. However, due to inertia, it takes some
time for the energy resources to re-enter a safe frequency region
until the power network eventually converges to steady state. Hence,
during transients, generators are still in danger of reaching their
frequency limits and being tripped, which may in turn cause
blackouts. This phenomenon tends to happen more frequently in modern
power networks due to low inertia and highly-dynamic units.
Therefore, there is a need to analyze the transient behavior of power
networks and design controllers that ensure the safe evolution of the
system.

\emph{Literature review.}  Transient stability refers to the ability
of power networks to maintain synchronism after being subjected to a
disturbance, see e.g.,~\citep{PK-JP:04}. Many works, see
e.g.,~\citep{HDC:11,FD-MC-FB:13,PJM-JH-JK-HJS:14}, provide conditions
to ensure synchronicity and investigate their relationship with the
topology of the power network. However, even if network synchronism
holds, system transient trajectory may enter unsafe regions, e.g.,
transient frequency may violate individual generator's frequency
limits, causing generator failure and leading to
blackouts~\citep{PK:94}. Hence, various techniques have been proposed
to improve transient behavior. These include resource re-dispatch with
transient stability constraints~\citep{AA-EBM:06,TTN-VLN-AK:11};
thyristor-controlled series capacitor compensation to optimize
transmission impedance and keep power transfer
constant~\citep{TG-JP:01}; the use of power system stabilizers to damp
out low frequency inter-machine
oscillations~\citep{MAM-HRP-MA-MJH:14}, and placing virtual inertia in
power networks to mitigate transient
effects~\citep{TSB-TL-DJH:15,BKP-SB-FD:17}. While these approaches
have a qualitative effect on transient behavior, they do not offer
strict guarantees as to whether the transient frequency stays within a
specific region. Furthermore, the approach by~\cite{TSB-TL-DJH:15}
requires a priori knowledge of the time evolution of the disturbance
trajectories and an estimation of the transient overshoot. Alternative
approaches rely on the idea of identifying the disturbances that may
cause undesirable transient behaviors using forward and backward
reachability analysis, see
e.g.,~\citep{MA:14,YCC-ADD:12,HC-PJS-SVD:16} and our previous
work~\citep{YZ-JC:17-acc}.  The lack of works that provide tools for
transient frequency control motivates us here to design feedback
controllers for the generators that guarantee simultaneously the
stability of the power network and the desired transient frequency
behavior. Our design is inspired by the controller-design approach to
safety-constrained systems taken by~\cite{ADA-XX-JWG-PT:16}, where the
safety region is encoded as the zero-sublevel set of a barrier
function and safety is ensured by constraining the evolution of the
function along the system trajectories.

\emph{Statement of contributions.}
The main result of the paper is the synthesis of a Lipschitz
continuous, distributed controller, available at specific individual
generator nodes, that satisfies the following requirements (i) renders
the closed-loop power network asymptotically stable; (ii) for each
controlled generator node, if its initial frequency belongs to a
desired safe frequency region, then its frequency trajectory stays in
it for all subsequent time; and (iii) if, instead, its initial
frequency does not belong to the safe region, then the frequency
trajectory enters it in finite time, and once there, never leaves.
Our technical approach to achieve this combines Lyapunov stability and
set invariance theory. We first show that requirement (iii)
automatically holds if (i) and (ii) hold true, and we thereby focus
our attention on the latter. For each one of these requirements, we
provide equivalent mathematical formulations that are amenable to
control design. Regarding (i), we consider an energy function for the
power system and formalize it as identifying a controller that
guarantees that the time evolution of this energy function along every
trajectory of the dynamics is non-decreasing. Regarding (ii), we show
that this condition is equivalent to having the controller make the
safe frequency interval forward invariant.  To avoid discontinuities
in the controller design on the boundary of the invariant set, we
report to the idea of barrier functions to have
the control effort gradually kick in as the state trajectory
approaches the boundary.  Our final step is to use the identified
constraints to synthesize a specific controller that satisfies both
and is distributed. The latter is a consequence of the fact that, for
each bus, the constraints only involve the state of the bus and that
of neighboring states.
We analyze its robustness properties against measure error and
parameter uncertainty, quantify its magnitude when the initial state
is uncertain, and provide an estimation on the frequency convergence
rate from the unsafe to the safe region for each controlled
generator. Finally, we illustrate the performance and design
trade-offs of the proposed controller on the IEEE 39-bus power
network.

\section{Preliminaries}\label{section:prelimiaries}
In this section we introduce basic notation and notions from set
invariance and graph theory.

\vspace*{-1.5ex}
\paragraph*{Notation.}
Let $\naturals$, $\real$, $\real_{>}$, and $\real_{\geqslant}$ denote
the set of natural, real, strictly positive, and nonnegative real
numbers, respectively.  Variables are assumed to belong to the
Euclidean space unless specified otherwise. For $a,b\in\naturals$,
denote $[a,b]_{\naturals}\triangleq\{x\in\naturals\ |\ a\leqslant
x\leqslant b\}$. Given $\mathcal{C} \subset \real^{n}$,
$\partial\mathcal{C}$ denotes its boundary.  We let $\|\cdot\|_{2}$
denote the 2-norm on $\real^{n}$. For a point $x\in\real^{n}$ and
$r\in\real_{>}$, denote
$B_{r}(x)\triangleq\setdef{x'\in\real^{n}}{\|x'-x\|_{2}\leqslant r}$.
Denote $\ones_n$ and $\zeros_n$ in $\real^n$ as the vector of all ones
and zeros, respectively.  For $A\in\mathbb{R}^{m\times n}$, let
$[A]_i$ and $[A]_{ij}$ denote its $i$th row and $(i,j)$th element. We
denote by $A^{\dagger}$ its unique Moore-Penrose pseudoinverse and by
$\range(A)$ its column space. A continuous function
$\alpha:\real\rightarrow \real$ is of class-$\mathcal{K}$ if it is
strictly increasing and $\alpha(0)=0$.  Given a differentiable
function $l:\real^{n}\rightarrow\real$, we let $\nabla l$ denote its
gradient. A function $f:\real_{\geqslant
}\times\real^{n}\rightarrow\real^{n},\ (t,x)\rightarrow f(t,x)$ is
Lipschitz in $x$ (uniformly in $t$)
%
%
if for every $x_{0}\in\real^{n}$, there exist $L,r>0$ such that
$\|f(t,x)-f(t,y)\|_{2}\leqslant L\|x-y\|_{2}$ for any $x,y\in
B_{r}(x_{0})$ and any $t\geqslant 0$.

\vspace*{-1.5ex}
\paragraph*{Set invariance.}
We introduce here notions of forward invariance~\cite{HKK:02}.
Consider the non-autonomous system on~$\real^{n}$,
\begin{align}\label{eqn:nonlinear}
  \dot x=f(t,x), \quad x(0)=x_{0},
\end{align}
where $f:\real_{\geqslant}\times\real^{n}\rightarrow\real^{n}$. We
assume $f$ is piecewise continuous in $t$ and Lipschitz in $x$, so
that the solution of~(\ref{eqn:nonlinear}) exists and is unique.  A
set $\mathcal{C}\in\real^{n}$ is \textit{(forward) invariant} for
system~\eqref{eqn:nonlinear} if for every initial condition $x_{0}\in
\mathcal{C}$, the solution starting from $x_0$ satisfies $x(t)\in
\mathcal{C}$ for all $t\geqslant 0$.  
The following result states a sufficient and necessary
condition for a set to be forward invariant for~\eqref{eqn:nonlinear}.

\begin{lemma}\longthmtitle{Nagumo's
    Theorem~\cite{FB-SM:08}}\label{lemma:Nagumo}
  Let $l:\real^{n}\rightarrow\real$ be continuously differentiable and
  let $ \mathcal{C}\triangleq\setdef{x}{l(x)\leqslant 0}$.  Suppose
  that for all $x\in\mathcal{C}$, there exists $s\in\real^{n}$ such
  that $l(x)+\nabla l(x)^{T}s<0$. Furthermore, suppose there exists a
  Lipschitz function $\phi:\real^{n}\rightarrow\real^{n}$ such that
  $\nabla l(x)^{T}\phi(x)<0$ for all $x\in\partial\mathcal{C}$.  Then
  $\mathcal{C}$ is forward invariant if and only if $\nabla
  l(x)^{T}f(t,x)\leqslant 0$ for all $x\in\partial\mathcal{C}$.
\end{lemma}

The assumptions in Nagumo's Theorem ensure that the set $\mathcal{C}$
is regular enough to have a well-defined interior and boundary.

\vspace*{-1.5ex}
\paragraph*{Graph theory.}
We present basic notions in algebraic graph theory
from~\cite{FB-JC-SM:08cor,NB:94}. An undirected graph is a pair
$\mathcal{G} = \mathcal(\mathcal{I},\mathcal{E})$, where $\mathcal{I}
= \{1,\dots,n\}$ is the vertex set and $\mathcal{E}=\{e_{1},\dots,
e_{m}\} \subseteq \mathcal{I} \times \mathcal{I}$ is the edge set.  A
path is an ordered sequence of vertices such that any pair of
consecutive vertices in the sequence is an edge of the graph. A graph
is connected if there exists a path between any two vertices. Two
nodes are neighbors if there exists an edge linking them. Denote by
$\mathcal{N}(i)$ the set of neighbors of node~$i$. For each edge
$e_{k} \in \mathcal{E}$ with vertices $i,j$, the orientation procedure
consists of choosing either $i$ or $j$ to be the positive end of
$e_{k}$ and the other vertex to be the negative end. The incidence
matrix $D=(d_{k i}) \in \mathbb{R}^{m \times n}$ associated with
$\mathcal{G}$ is then defined as
\begin{align*}
  d_{k i} =
  \begin{cases}
    1 & \text{if $i$ is the positive end of $e_{k}$},
    \\
    - 1 & \text{if $i$ is the negative end of $e_{k}$},
    \\
    0 & \text{otherwise}.
  \end{cases}
\end{align*}

\section{Problem statement}\label{section:problem-statement}
In this section we introduce the dynamical model for the power network
and state our control objective.

\subsection{Power network model}
The power network is encoded by a connected undirected graph
$\mathcal{G} = (\mathcal{I},\mathcal{E})$, where $\mathcal{I} =
\{1,2,\cdots,n\}$ is the collection of buses and $\mathcal{E} =
\{e_{1},\cdots,e_{m}\}\subseteq\mathcal{I}\times\mathcal{I}$ is the
collection of transmission lines.  For each node $i\in\mathcal{I}$,
let $\theta_{i}\in\real$, $\omega_{i}\in\real$ and $p_{i}\in\real$
denote its voltage angle, shifted voltage frequency relative to the
nominal frequency, 
and constant active power injection, respectively.  We partition buses
into $\CC$ and $\mathcal{I} \backslash \CC$, where every bus $i\in\CC$
requires an individual transient frequency regulation realized via an
exogenous control command~$u_{i}$. The dynamics is described by the
swing equations for voltage angles and frequencies,
\begin{align}\label{eqn:swing-equations-dynamics}
  \dot\theta_{i}(t) &\hspace{-0.05cm} =\omega_{i}(t), \ \forall
  i\in\mathcal{I},
  \\
  M_{i}\dot\omega_{i}(t) & \hspace{-0.05cm}=\hspace{-0.05cm}
  -E_{i}\omega_{i}(t) - \hspace*{-2ex} \sum_{j\in\mathcal{N}(i)}
  \hspace*{-1.5ex} b_{ij} \sin(\theta_{i}(t)-\theta_{j}(t))
  \hspace{-0.05cm}+u_{i}(t)\hspace{-0.05cm}+ p_{i},\
  \hspace{-0.05cm} \forall i\in\CC,\notag
  \\
  M_{i}\dot\omega_{i}(t) & \hspace{-0.05cm}=\hspace{-0.05cm}
  -E_{i}\omega_{i}(t) - \hspace*{-2ex} \sum_{j\in\mathcal{N}(i)}
  \hspace*{-1.5ex} b_{ij} \sin(\theta_{i}(t)-\theta_{j}(t))
  \hspace{-0.05cm}+ p_{i},\ \hspace{-0.05cm}\forall
  i\in\mathcal{I}\backslash\CC, \notag
\end{align}
where $b_{ij}\in\real_{>}$ is the susceptance of the line connecting
bus $i$ and~$j$, and $M_{i} \in \real_{\geqslant}$ and $E_{i} \in
\real_{\geqslant}$ are the inertia and damping coefficients of bus $i
\in \mathcal{I}$.  For simplicity, we assume that they are all
strictly positive.

For our purposes, it is convenient to rewrite the
dynamics~\eqref{eqn:swing-equations-dynamics} in a more compact way.
Let $\theta \triangleq [\theta_{1}, \cdots, \theta_{n}]^{T} \in
\real^{n}$, $\omega \triangleq [\omega_{1}, \cdots, \theta_{n}]^{T}
\in \real^{n}$ and $p\triangleq [p_{1},\cdots,p_{n}]^{T}\in\real^{n}$
be the collection of voltage angles, frequencies, and power
injections.  Let $D\in\real^{m\times n}$ be the incidence matrix
corresponding to an arbitrary graph orientation, and define the
voltage angle difference vector
\begin{align}\label{eqn:state-transformation}
  \lambda \triangleq D\theta \in\real^{m} .
\end{align}
Denote by $Y_{b}\in\real^{m\times m}$ the diagonal matrix whose $k$th
diagonal item represents the susceptance of the transmission line
$e_{k}$ connecting bus $i$ and $j$, i.e., $[Y_{b}]_{k,k}=b_{ij},$ for
$k=1,2,\cdots, m$.  We re-write the
dynamics~\eqref{eqn:swing-equations-dynamics} in terms of $\lambda$
and $\omega$ as
\begin{subequations}\label{eqn:dynamics-2}
  \begin{align}
    \dot \lambda (t)
    &= D\omega(t),\label{eqn:dynamics-2a}
    \\
    M_{i}\dot\omega_{i}(t) 
    &=
      -E_{i}\omega_{i}(t)-[D^{T}Y_{b}]_{i}\sin\lambda(t)+u_{i}(t)+p_{i},
      \ \forall i\in\CC,\label{eqn:dynamics-2b}
    \\
    M_{i}\dot\omega_{i}(t)
    &=-E_{i}\omega_{i}(t)-[D^{T}Y_{b}]_{i}\sin\lambda(t)+p_{i},
      \ \forall
      i\in\mathcal{I}\backslash\CC,\label{eqn:dynamics-2c}
  \end{align}
\end{subequations}
where $\sin\lambda(t)\in\real^{m}$ is the component-wise sine
value of $\lambda(t)$. Note that the
transformation~\eqref{eqn:state-transformation} enforces
$  \lambda(0) \in \range(D)$.
We refer to an initial condition satisfying this equation as
\emph{admissible}.  When convenient, for conciseness, we use $
x(t)\triangleq \left(\lambda(t),\omega(t)\right)\in\real^{m+n}
$
to denote the collection of all states, and we neglect its dependence
on $t$ if the context is clear.

The trajectories $(\lambda(t),\omega(t))$ locally converge to a unique
equilibrium point if all $u_{i}$'s are set to zero. Specifically,
let $L\triangleq D^{T}Y_{b}D$ and $L^{\dagger}$ be its pseudoinverse.
Define $\omega^{\infty}\triangleq\frac{\sum_{i=1}^{n}
  p_{i}}{\sum_{i=1}^{n}E_{i}}$,
$E\triangleq\text{diag}(E_{1},E_{2},\cdots,E_{n})$, and $\tilde
p\triangleq p-\omega^{\infty}E\ones_{n}$.  If
\begin{align}\label{ineq:sufficient-eq}
  \|L^{\dagger}\tilde p\|_{\mathcal{E},\infty}<1,
\end{align}
where $\|y\|_{\mathcal{E},\infty} \triangleq
\max_{(i,j)\in\mathcal{E}} |y_{i}-y_{j}|$, then there exists
$\lambda^{\infty}\in \Gamma
\triangleq\setdef{\lambda}{|\lambda_{i}|<\pi/2}$ unique in
$\Gamma\closure\triangleq\setdef{\lambda}{|\lambda_{i}|\leqslant\pi/2}$ such that
\begin{align}\label{eqn:lambda-solution}
  \tilde p = D^{T}Y_{b}\sin \lambda^{\infty} \text{ and }
  \lambda^{\infty} \in \range (D).
\end{align}
According to \cite[Lemma 2 and inequality (S17)]{FD-MC-FB:13},
system~\eqref{eqn:dynamics-2} with $u_{i}\equiv0$ for every
$i\in\CC$, $(\lambda^{\infty},\omega^{\infty}\ones_{n})$ is
stable. Furthermore, $(\lambda(t),\omega(t))$ locally converges to
$(\lambda^{\infty},\omega^{\infty}\ones_{n})$ provided $\lambda(0) \in
\range(D)$. Throughout the rest of the paper, we assume that
condition~\eqref{ineq:sufficient-eq} holds.

\subsection{Control goal}
Our goal is to design a state-feedback controller for each bus
$i\in\CC$ that guarantees that the frequency transient behavior stays
within desired safety bounds while, at the same time, preserving the
stability properties that the system~\eqref{eqn:dynamics-2} enjoys
when no external input $u_{i}$ is present.  We state these
requirements explicitly next.

%
%

\emph{Stability and convergence requirement:} Since the
system~\eqref{eqn:dynamics-2} without $u_{i}$ is locally stable, we
require that the same system with the proposed controller $u_{i}$ is
also locally stable. Furthermore, for every admissible initial
condition, the two systems should converge to the same equilibrium
$(\lambda^{\infty},\omega^{\infty}\ones_{n})$, meaning that $u_{i}$
only affects the transient behavior.

\emph{Frequency invariance requirement:} For each $i\in\CC$, let
$\underline\omega_{i}\in\real$ and $\bar\omega_{i}\in\real$ be lower
and upper safe frequency bounds, where
$\underline\omega_{i}<\bar\omega_{i}$.  We require that the frequency
$\omega_{i}(t)$ stays inside the safe region
$[\underline\omega_{i},\bar\omega_{i}]$ for any $t> 0$, provided that
the initial frequency $\omega_{i}(0)$ lies inside
$[\underline\omega_{i},\bar\omega_{i}]$.  This forward invariance
requirement corresponds to underfrequency/overfrequency avoidance.

\emph{Attractivity requirement:} If, for some $i\in\CC$, the initial
frequency $\omega_{i}(0)\notin[\underline\omega_{i},\bar\omega_{i}]$,
then after a finite time, $\omega_{i}$ enters the safe region and
never leaves afterwards. This requirement corresponds to
underfrequency/overfrequency recovery.

In addition to these requirements, we also seek the designed
controller to be Lipschitz as a function of the state. This guarantees
the existence and uniqueness of solutions for the closed-loop system
and, at the same time, provides robustness for practical
implementation against errors in state measurements.

\begin{remark}\longthmtitle{Selection of buses with transient
    frequency specification}\label{rmk:selction-node}
  {\rm The set $\CC$ consists of buses belonging to either of the
    following two types: a) buses with specified over/underfrequency
    requirement~\citep{PP-PSK-CWT:06} and b) buses whose transient
    frequency behavior is key in evaluating system performance, or are
    used as indexes for load shedding schemes~\citep{NWM-KC-MS:11}. We
    assume each individual bus in $\CC$ is equipped with an external
    input directly tuning its transient behavior. We show later that
    this is necessary condition to obtain frequency invariance
    guarantees.} \oprocend
\end{remark}

Note that the attractivity requirement is automatically satisfied once
the controller meets the first two requirements, provided that
$\omega^{\infty}\in (\underline\omega_{i},\bar\omega_{i})$. However,
in general it is still of interest to provide estimates for how fast
the frequency reaches the safe region.  Our objective is to design a
controller that satisfies the above three requirements simultaneously
and is distributed, in the sense that each bus can implement it using
its own information and that of its neighboring buses and transmission
lines.


\section{Constraints on controller design}

In this section, we identify constraints on the controller design that
provide sufficient conditions to ensure, on the one hand, the
stability and convergence requirement and, on the other hand, the
frequency invariance requirement.

\subsection{Constraint ensuring stability and convergence}
We establish a stability constraint by identifying an energy function
and restricting the input so that its evolution along every trajectory
of the closed-loop dynamics is monotonically non-increasing.  We
select the following energy function~\citep{TLV-HDN-AM-JS-KT:17}
\begin{align}\label{eqn:energy-func}
  V(\lambda,\omega)\triangleq\frac{1}{2}\sum_{i=1}^{n}M_{i}
  (\omega_{i}-\omega^{\infty})^{2} +
  \sum_{j=1}^{m}[Y_{b}]_{j,j}a(\lambda_{j}),
\end{align}
where $a(\lambda_{j}) \triangleq \cos\lambda_{j}^{\infty} -
\cos\lambda_{j} - \lambda_{j}\sin\lambda_{j}^{\infty} +
\lambda_{j}^{\infty}\sin\lambda_{j}^{\infty}$.  The next result uses
the LaSalle Invariance Principle to show this property.

\begin{lemma}\longthmtitle{Sufficient condition for local stability
    and convergence}\label{lemma:sufficient-stability-convergence}
  Consider the system~\eqref{eqn:dynamics-2}.  Under
  condition~\eqref{ineq:sufficient-eq}, further suppose that, for
  every $i\in\CC$, $u_{i}:\real^{m+n}\times\real^{n}\rightarrow\real,\
  (x,y)\mapsto u_{i}(x,y)$ is Lipschitz in $x$. Let $c \triangleq
  \min_{ \lambda\in\partial\Gamma
    \closure}V(\lambda,\omega^{\infty}\ones_{n})$ and define
  \begin{align}\label{set:region}
    \Phi\triangleq\setdef{(\lambda,\omega)}{\lambda\in
      \Gamma\closure,\ V(\lambda,\omega)\leqslant c/\beta} 
  \end{align}
  with $\beta\in\real_{>}$. If for every $i\in\CC$, $x\in\real^{m+n}$,
  and $p\in\real^{n}$,
  \begin{subequations}\label{ineq:stabilize-constraints-2}
    \begin{align}
      (\omega_{i}-\omega^{\infty})u_{i}(x,p) &\leqslant 0 \quad
      \text{if }
      \omega_{i}\neq\omega^{\infty},\label{ineq:stabilize-constraints-2a}
      \\
      u_{i}(x,p) &=0 \quad \text{if
      }\omega_{i}=\omega^{\infty},\label{ineq:stabilize-constraints-2b}
    \end{align}
  \end{subequations}
  then the following results hold provided $\lambda(0) \in \range (D)$
  and $(\lambda(0),\omega(0))\in\Phi$ for some $\beta>1$:
  \begin{enumerate}
  \item \label{item:solution} The solution of the closed-loop system
    exists and is unique for any $t\geqslant 0$;
  \item\label{item:invariance} $\lambda(t)\in \range(D)$ and
    $(\lambda(t),\omega(t))\in\Phi$ for any $t\geqslant 0$;
  \item\label{item:convergence-stability}
    $(\lambda^{\infty},\omega^{\infty}\ones_{n})$ is stable, and
    $(\lambda(t),\omega(t))\rightarrow(\lambda^{\infty},
    \omega^{\infty}\ones_{n})$ as $t\rightarrow \infty$.
  \end{enumerate}
\end{lemma}
\begin{pf}
  To prove\emph{~\ref{item:solution}}, as $(x,y) \mapsto u_{i}(x,y)$
  is Lipschitz in~$x$, there exists a unique local solution over
  $[0,\delta]$ for some $\delta>0$, according
  to~\cite[Theorem~3.1]{HKK:02}. Let $[0,T)$ be the maximal interval
  of existence. We then show that $\Phi$ is non-empty and compact, and
  that $(\lambda(t),\omega(t))$ lies entirely in $\Phi$ for any
  $t\in[0,T)$. These two facts together,
  by~\cite[Theorem~3.3]{HKK:02}, imply the existence and uniqueness of
  the solution for every $t\geqslant 0$.  To show the non-emptiness of
  $\Phi$, note that in~\eqref{eqn:energy-func} if
  $|\lambda_{i}|\leqslant \pi/2$ and $|\lambda_{i}^{\infty}|<\pi/2$,
  then $a(\lambda_{i})\geqslant 0$, which implies that
  $V(\lambda,\omega)\geqslant 0$ for every $\lambda\in\Gamma\closure$
  and every $ \omega\in\real^{n}$; hence $c\geqslant 0$.  Then
  $(\lambda^{\infty},\omega^{\infty}\ones_{n})\in\Phi$ as
  $V(\lambda^{\infty},\omega^{\infty}\ones_{n})=0$.
  %
  To show the compactness of $\Phi$, note that the set is clearly
  closed. Since the polytope $\Gamma_{\closure}$ is bounded, the variable
  $\lambda$ is bounded too.  Therefore,
  $a(\lambda_{i})$ is bounded for every
  $i\in[1,m]_\naturals$. Since $ V(\lambda,\omega) \le c/\beta$, we
  deduce that $\sum_{i=1}^{n}M_{i}(\omega_{i}-\omega^{\infty})^{2}$ is
  bounded, implying that $\omega$ is bounded. Hence, $\Phi$ is
  bounded.

  Regarding statement\emph{~\ref{item:invariance}}, note that
  $\lambda(t)\in \range(D)$ holds for every $t\geqslant 0$ since both
  $\lambda(0)$ and $\dot\lambda(t)$ lie in $\range(D)$. To establish
  the invariance of $\Phi$, we examine the evolution of the function
  $V$ along the dynamics~\eqref{eqn:dynamics-2},
  \begin{align*}
    \dot V(\lambda,\omega)
    &=\sum_{i=1}^{n}(\omega_{i}-\omega^{\infty})
      \left(-E_{i}\omega_{i}-[D^{T}Y_{b}]_{i}\sin\lambda+p_{i}\right)
    \\
    &+\sum_{i \in \CC}(\omega_{i}-\omega^{\infty})u_{i}(x,p) +
      \sum_{j=1}^{m}[Y_{b}]_{j,j}(\sin\lambda_{j} -
      \sin\lambda_{j}^{\infty})[D]_{j}\omega  
    \\
    &=-\sum_{i=1}^{n}E_{i}(\omega_{i}-\omega^{\infty})^{2} +
      \sum_{i\in\CC}(\omega_{i}-\omega^{\infty})u_{i}(x,p)
    \\
    &\leqslant
      -\sum_{i=1}^{m}E_{i}(\omega_{i}-\omega^{\infty})^{2}\leqslant 0,
  \end{align*}
  where we have employed~\eqref{eqn:derivative-step-0} in the second equality.
  \begin{figure*}[htb!]
    \begin{align}
      &\sum_{i=1}^{n}(\omega_{i}-\omega^{\infty})
        \left(-[D^{T}Y_{b}]_{i}\sin\lambda+p_{i}-\omega^{\infty} E_i\right) +
        \sum_{j=1}^{m}[Y_{b}]_{j,j}(\sin\lambda_{j}-\sin\lambda_{j}^{\infty})[D]_{j}
        \omega\notag  
      \\
      =&\sum_{i=1}^{n}(\omega_{i}-\omega^{\infty})
         \left(-[D^{T}Y_{b}]_{i}\sin\lambda+p_{i}-\omega^{\infty} E_i\right) +
         \sum_{j=1}^{m}(\sin\lambda_{j} -
         \sin\lambda_{j}^{\infty})[Y_{b}D]_{j}(\omega-\omega^{\infty}\ones_{n})\notag
      \\
      =&\sum_{i=1}^{n}(\omega_{i}-\omega^{\infty} )
         \left(p_{i}-\omega^{\infty} E_i\right) -
         \sum_{j=1}^{m}(\sin\lambda_{j}^{\infty})[Y_{b}D]_{j}(\omega-\omega^{\infty}
         \ones_{n})  
         \notag
      \\
      =&\sum_{i=1}^{n}(\omega_{i}-\omega^{\infty} )
         \left(p_{i}-\omega^{\infty}E_i-D^{T}Y_{b}\sin\lambda_{i}^{\infty}\right)
         =(\omega-\omega^{\infty}\ones_{n})^{T}(\tilde
         p-D^{T}Y_{b}\sin\lambda^{\infty})=0.\label{eqn:derivative-step-0}
    \end{align}
    \hrulefill
  \end{figure*}
  This monotonicity of $V$ implies that the constraint
  $V(\lambda,\omega)\leqslant c/\beta$ defining $\Phi$ can never be
  violated.  Now if there exists a time $t_{1}>0$ such that
  $(\lambda(t_{1}),\omega(t_{1}))\notin\Phi$, then it must be the case
  where $\lambda(t_{1})\notin\Gamma$. By the continuity of the
  trajectory, there must exist another time $t_{2}$ before $t_{1}$
  such that $\lambda(t_{2})\in\partial\Gamma_{\closure}$, in which
  case $V(\lambda(t_{2}),\omega(t_{2}))\geqslant
  V(\lambda(t_{2}),\omega^{\infty}\ones_{n})\geqslant c>c/\beta$,
  which is a contradiction. Hence $\Phi$ is invariant.
  
  To prove~\emph{\ref{item:convergence-stability}}, notice that, for
  any $ (\lambda,\omega)\in\Phi,\ \dot V(\lambda,\omega)\leqslant 0$;
  second, $V(\lambda^{\infty},\omega^{\infty}\ones_{n})=0$; third,
  $V(\lambda,\omega)>0,$ for every $(\lambda,\omega)\in\Phi$ with
  $(\lambda,\omega)\neq(\lambda^{\infty},\omega^{\infty}\ones_{n})$.
  By~\cite[Theorem~4.1]{HKK:02},
  $(\lambda^{\infty},\omega^{\infty}\ones_{n})$ is stable.  Finally,
  to establish convergence, let
  \begin{align}\label{set:omega}
    \Omega \triangleq \Phi \cap \setdef{(\lambda,\omega)}{ \text{}
    \lambda \in \range(D)}.
  \end{align} 
  Note that $(\lambda(0),\omega(0))\in\Omega$. Clearly, the set
  $\Omega$ is compact and invariant with respect to the
  dynamics~(\ref{eqn:dynamics-2a})-(\ref{eqn:dynamics-2c}) with
  controller satisfying~(\ref{ineq:stabilize-constraints-2}). Noticing
  that $\dot V(\lambda,\omega)=0$ implies
  $\omega=\omega^{\infty}\ones_{n}$, let
  $S\triangleq 
  \setdef{(\lambda,\omega)}{\omega=\omega^{\infty}\ones_{n}}\bigcap\Omega$.
  It is easy to see that no solution can identically stay in $S$ other
  than the trivial solution
  $(\lambda(t),\omega(t))\equiv(\lambda^{\infty},\omega^{\infty}\ones_{n})$. The
  conclusion then follows from the LaSalle Invariance
  Principle~\cite[Theorem~4.4]{HKK:02}. \qed
\end{pf}

\begin{remark}\longthmtitle{Computation of the region of
    attraction}\label{rmk:app-invairance}
  {\rm The set~$\Phi$ is an estimate of the region of attraction but
    its explicit computation requires the solution of a non-convex
    optimization problem to determine the value of~$c$.  We can
    equivalently compute $c$ by solving $2m$ convex problems. For each
    $j\in[1,m]_{\naturals}$, let
    \begin{align*}
      \bar c_{j} \triangleq \hspace*{-2ex} \min_{
      \begin{subarray}{c}
        \lambda_{j}=\pi/2
        \\
        |\lambda_{i}|\leqslant \pi/2, \, \forall i \neq j
      \end{subarray}
      }
      V( \lambda,\omega^{\infty}\ones_{n}),
      \quad
      \underline c_{j}
      \triangleq \hspace*{-2ex} \min_{
      \begin{subarray}{c}
        \lambda_{j}=-\pi/2
        \\
        |\lambda_{i}|\leqslant \pi/2, \, \forall i \neq j
      \end{subarray}
      }
      V( \lambda,\omega^{\infty}\ones_{n}).
    \end{align*}
    Note that these problems are convex, as the Hessian of
    $V(\tilde\lambda, \omega^{\infty}\ones_{n})$ with respect to
    $\tilde\lambda$, $\nabla^{2}V = \diag([Y_{b}]_{1,1}
    \cos(\lambda_{1}), \cdots,$\\$[Y_{b}]_{m,m}\cos(\lambda_{m}))$, is
    positive definite on $\Gamma_{\closure}$, and the feasible set is
    a closed convex subset of $\Gamma_{\closure}$.  One can easily see
    that $c=\min_{j\in[1,m]_{\naturals}}\{\bar c_{j},\underline
    c_{j}\}$.

    On the other hand, although it is easy to check if a given initial
    state belongs to $\Phi$, it is difficult to characterize its
    geometric shape. The work~\citep{TLV-HDN-AM-JS-KT:18} shows that,
    for suitable $\bar c>0$ determined via a convex quadratic program,
    the ellipsoid
    \begin{align*}
      \bar\Phi\triangleq \setdef{(\lambda,\omega)}{\bar
      V(\omega,\lambda)\leqslant \bar c}
    \end{align*}
    is a subset of $\Phi$ (here $\bar V(\omega,\lambda) \triangleq
    \frac{1}{2}\sum_{i=1}^{n}M_{i} (\omega_{i}-\omega^{\infty})^{2} +
    \frac{1}{2} \sum_{j=1}^{m}[Y_{b}]_{j,j}
    (\lambda_{j}-\lambda_{j}^{\infty})^{2}$ is quadratic).
    Lemma~\ref{lemma:sufficient-stability-convergence} remains valid
    if $\Phi$ is replaced by~$\bar\Phi$. } \oprocend
\end{remark}

\subsection{Constraint ensuring frequency
  invariance}\label{sec:constraint-freq}

We next focus our attention on the frequency invariance
requirement. We start by defining the invariant sets we are interested
in,
\begin{align}\label{eqn:barrier-function-unsymmetric}
  \bar{\mathcal{C}}_{i}\triangleq\setdef{x}{\omega_{i}-\bar\omega_{i}\leqslant
    0}, \quad \underline{\mathcal{C}}_{i} \triangleq\setdef{x}{
    \underline\omega_{i}-\omega_{i}\leqslant 0}.
\end{align}
The characterization stated in the next result directly follows from
Nagumo's Theorem.

\begin{lemma}\longthmtitle{Sufficient and necessary condition for
    frequency invariance}\label{lemma:frequency-invariance}
  Assume that the solution of~\eqref{eqn:dynamics-2} exists and is
  unique for every admissible initial condition. Then, for any
  $i\in\CC$, the sets $\bar{\mathcal{C}}_{i}$ and
  $\underline{\mathcal{C}}_{i}$ are invariant if and only if for every
  $x\in\real^{m+n}$ and $p\in\real^{n}$,
  \begin{subequations}\label{ineq:invariance-condition-unsymmetric-1}
    \begin{align}
      u_{i}(x,p)-q_{i}(x,p)\leqslant 0 \quad \text{if }
      \omega_{i}=\bar\omega_{i},\label{ineq:invariance-condition-unsymmetric-1a}
      \\
      -u_{i}(x,p)+q_{i}(x,p)\leqslant 0 \quad \text{if
      }\omega_{i}=\underline
      \omega_{i},\label{ineq:invariance-condition-unsymmetric-1b}
    \end{align}
  \end{subequations}
  where $ q_{i}(x,p)\triangleq
  E_{i}\omega_{i}+[D^{T}Y_{b}]_{i}\sin\lambda-p_{i}$.
\end{lemma}
\begin{pf}
  For simplicity, we only deal with the case of
  $\bar{\mathcal{C}}_{i}$ (the other case follows similarly).  For
  each $i \in \CC$, let $\bar l_{i}, \underline l_{i}: \real^n
  \rightarrow \real$ be defined by $\bar l_{i}(x)
  \triangleq\omega_{i}-\bar\omega_{i}$ and $\underline
  l_{i}(x)\triangleq-\omega_{i}+\underline\omega_{i}$.  Notice that,
  by letting $s=-\ones_{m+n}$ and $\phi(x)\equiv-\ones_{m+n}$, one has
  that $\bar l_{i}(x)+\nabla \bar l_{i}(x)^{T}s<0$ for every
  $x\in\bar{\mathcal{C}}_{i}$ and $\nabla \bar l_{i}(x)^{T}\phi(x)<0$
  for every $x\in\partial\bar{\mathcal{C}}_{i}$, and hence the
  assumptions in Nagumo's Theorem hold.  Denote by $f(t,x)$ the
  right-hand side of the dynamics~\eqref{eqn:dynamics-2}. Then
  $\bar{\mathcal{C}}_{i}$ is invariant if and only if $\nabla\bar
  l_{i}(x)^{T}f(t,x)\leqslant 0$ when $\omega_{i}(t)=\bar\omega_{i}$,
  which is equivalent
  to~\eqref{ineq:invariance-condition-unsymmetric-1a}. \qed
\end{pf}

From Lemma~\ref{lemma:frequency-invariance}, one sees that if some bus
$j\in\CC$ does not possess an external control input (i.e.,
$u_{j}\equiv0$), then one can not guarantee the invariance of
$\bar{\mathcal{C}}_{j}$ and $\underline{\mathcal{C}}_{j}$, since
without an active control signal,
condition~\eqref{ineq:invariance-condition-unsymmetric-1} can easily
be violated.
The characterization of Lemma~\ref{lemma:frequency-invariance} points
to the value of the input at the boundary of $\bar{\mathcal{C}}_{i}$
and $\underline{\mathcal{C}}_{i}$. However, having a controller that
is only nonvanishing at such points is undesirable, as the actuator
effort would be discontinuous, affecting the system evolution. A more
sensible policy is to have the controller become active as the system
state gets closer to the boundary of these sets, and do so in a
gradual way. This is captured by the following result.

\begin{lemma}\longthmtitle{Sufficient condition for frequency
    invariance}\label{lemma:sufficent-frequecy-invariance}
  Assume that the solution of~\eqref{eqn:dynamics-2} exists and is
  unique for every admissible initial condition.  For each $i\in\CC$,
  let $\bar\omega_{i}^{\operatorname{th}},\
  \underline\omega_{i}^{\operatorname{th}}\in\real$ be such that
  $\underline\omega_{i}<\underline\omega_{i}^{\operatorname{th}} <
  \bar\omega_{i}^{\operatorname{th}}<\bar\omega_{i}$ and let $\bar\alpha_{i}$
  and $\underline\alpha_{i}$ be functions of class-$\mathcal{K}$.  If
  for every $x\in\real^{m+n}$ and $p\in\real^{n}$,
  \begin{subequations}\label{ineq:invariance-condition-unsymmetric-4}
    \begin{align}
      (\omega_{i}-\bar\omega_{i}^{\operatorname{th}})(u_{i}(x,p)-q_{i}(x,p))\leqslant
      -\bar\alpha_{i}(\omega_{i}-\bar\omega_{i}),
      \label{ineq:invariance-condition-unsymmetric-3-a}
      \end{align}
      if $\bar\omega_{i}^{\operatorname{th}}<\omega_{i}\leqslant
      \bar\omega_{i}$, and 
      \begin{align}
        (\underline\omega_{i}^{\operatorname{th}}-\omega_{i})(-u_{i}(x,p)+q_{i}(x,p))\leqslant
        -\underline\alpha_{i}(\underline\omega_{i}-\omega_{i}),
        \label{ineq:invariance-condition-unsymmetric-3-b}
    \end{align}
  \end{subequations}
  if $ \underline\omega_{i}\leqslant \omega_{i}<
  \underline\omega_{i}^{\operatorname{th}}$, then $\bar{\mathcal{C}}_{i}$ and
  $\underline{\mathcal{C}}_{i}$ are invariant.
\end{lemma}

The proof of Lemma~\ref{lemma:sufficent-frequecy-invariance} follows
by noting that, when $\omega_{i}=\bar\omega_{i}$
(resp. $\omega_{i}=\underline\omega_{i}$),
condition~(\ref{ineq:invariance-condition-unsymmetric-3-a})
(resp.~(\ref{ineq:invariance-condition-unsymmetric-3-b}))
becomes~(\ref{ineq:invariance-condition-unsymmetric-1a})
(resp.~(\ref{ineq:invariance-condition-unsymmetric-1b})).  The
introduction of class-$\mathcal{K}$ functions enables the design of
controllers that gradually kick in as the margin for satisfying the
requirement for frequency invariance gets increasingly small. In fact,
using~\eqref{eqn:dynamics-2}, we can equivalently
write~\eqref{ineq:invariance-condition-unsymmetric-3-a} as
\begin{align}\label{ineq:derivative-class-K}
  M\dot\omega_{i}\leqslant -
  \bar\alpha_{i}(\omega_{i}-\bar\omega_{i})/(\omega_{i} -
  \bar\omega_{i}^{\operatorname{th}}), \quad \text{if }
  \bar\omega_{i}^{\operatorname{th}}<\omega_{i}\leqslant \bar\omega_{i}.
\end{align}
Notice that, as $\omega_{i}$ grows from the threshold
$\bar\omega_{i}^{\operatorname{th}}$ to the safe bound $\bar\omega_{i}$, the
value of $-\bar\alpha_{i} (\omega_{i}-\bar\omega_{i})/(\omega_{i} -
\bar\omega_{i}^{\operatorname{th}})$ monotonically decreases to 0.  Thus, the
constraint on $\dot\omega_{i}$ becomes tighter (while allowing
$\dot\omega_{i}$ to still be positive) as $\omega_{i}$ approaches
$\bar\omega_{i}$, and when $\omega_{i}$ hits $\bar\omega_{i}$,
prescribes $\dot\omega_{i}$ to be nonpositive to ensure invariance.
It is interesting to point out the trade-offs present in the choice of
class-$\mathcal{K}$ functions. A function with a large
derivative, for instance, corresponds to a controller design that
allows the derivative above to be significant near the boundary, at
the risk of increasing the sensitivity to changes in the state. We
re-examine this point later after introducing our specific controller
design.

\section{Distributed controller synthesis}
In this section we introduce a distributed controller design that
meets the stability and convergence
condition~\eqref{ineq:stabilize-constraints-2} as well as the
frequency invariance
condition~\eqref{ineq:invariance-condition-unsymmetric-4}.  Our next
result formally introduces this controller and characterizes its
continuity property.

\begin{proposition}\longthmtitle{Distributed frequency
    controller}\label{prop:L}
For each
  $i \in \CC$, let $\bar\alpha_{i}$ and $\underline\alpha_{i}$
  be Lipschitz functions of class-$\mathcal{K}$. Then,
  \begin{align}\label{eqn:stability-transient-controller-Lipschitz-4}
    u_{i}(x,p) \!=\!
    \begin{cases}
      \min\{0,\frac{-\bar\alpha_{i}(\omega_{i}-\bar\omega_{i})}{\omega_{i}
        - \bar\omega_{i}^{\operatorname{th}}}+q_{i}(x,p)\} &
      \omega_{i}>\bar\omega_{i}^{\operatorname{th}},
      \\
      0 & \underline\omega_{i}^{\operatorname{th}}\leqslant
      \omega_{i}\leqslant \bar\omega_{i}^{\operatorname{th}},
      \\
      \max\{0,\frac{\underline\alpha_{i}(\underline\omega_{i} -
        \omega_{i})}{\underline\omega_{i}^{\operatorname{th}}-\omega_{i}} +
      q_{i}(x,p)\} & \omega_{i}<\underline\omega_{i}^{\operatorname{th}},
    \end{cases}
  \end{align}
  is Lipschitz in its first argument.
\end{proposition}
\begin{pf}
  Let $i \in \CC$.  We show that for any $x\in\real^{m+n}$, there
  exist $L,r\in\real_{>}$ such that $|u_{i}(y,p)-u_{i}(z,p)|\leqslant
  L \|y-z\|$ for any $y,z\in B_{r}(x)$.  Notice that this condition
  holds true for $x$ belonging to $\mathbb{H} \triangleq
  \setdef{x\in\real^{m+n}}{\omega_{i}\neq\bar\omega_{i}^{\operatorname{th}},\
    \omega_{i}\neq\underline\omega_{i}^{\operatorname{th}}}$, in that
  $x\mapsto\frac{-\bar\alpha_{i}(\omega_{i} -
    \bar\omega_{i}))}{(\omega_{i}-\bar\omega_{i}^{\operatorname{th}})}+q_{i}(x,p)$
  (resp.  $x \mapsto \frac{\underline\alpha_{i}
    (\underline\omega_{i}-\omega_{i})}{\underline\omega_{i}^{\operatorname{th}}-\omega_{i}}
  + q_{i}(x,p)$) is Lipschitz for any $x$ in $\mathbb{H}$, and the
  $\min$ (resp. $\max$)) operator preserves Lipschitz
  continuity. Hence we only need to establish Lipschitzness for
  $x\nin\mathbb{H}$.  For simplicity we only reason for the case when
  $x$ satisfies $\omega_{i}=\bar\omega_{i}^{\operatorname{th}}$.
  Denote $ r_{0}\triangleq \min\{ \tfrac{1}{2}(\bar\omega_{i} -
  \bar\omega_{i}^{\operatorname{th}}),\ \tfrac{1}{2}
  (\bar\omega^{\operatorname{th}}_{i} -
  \underline\omega_{i}^{\operatorname{th}}) \}\in\real_{>}$. One can
  see that for any $x'\in B_{r_{0}}(x)$, it holds that
  $\underline\omega_{i}^{\operatorname{th}}\leqslant \omega_{i}$.
  Next we show that there always exists $r\leqslant r_{0}$ such that
  \begin{align}
    \frac{-\bar\alpha_{i}(\omega_{i}
      -\bar\omega_{i})}{(\omega_{i}-\bar\omega_{i}^{\operatorname{th}})} +
    q_{i}(x',p)>0, \label{eq:auxx}
  \end{align}
  for all $ x'\in B_{r}(x) \cap
  \setdef{x'}{\omega_{i}>\bar\omega_{i}^{\operatorname{th}}}$.  Notice that
  for any $x'\in B_{r}(x)$, $\omega_{i}-\bar\omega_{i}\leqslant
  \bar\omega_{i}^{\operatorname{th}} +
  r-\bar\omega_{i}\leqslant\omega_{i}^{\operatorname{th}} +
  (\bar\omega_{i}-\bar\omega_{i}^{\operatorname{th}})/2-\bar\omega_{i} =
  -(\bar\omega_{i}-\bar\omega_{i}^{\operatorname{th}})/2<0$, and
  $q_{i}(x',p)=\omega_{i}+[D^{T}]_{i}\lambda-p_{i}\geqslant -
  (n+1)\|x'\|_{2}-|p_{i}|$.  Therefore, it holds that
  \begin{align*}
    \frac{-\bar\alpha_{i}(\omega_{i} -
      \bar\omega_{i})}{(\omega_{i}-\bar\omega_{i}^{\operatorname{th}})} +
    q_{i}(x',p)\geqslant
    \frac{-\bar\alpha_{i}(\omega_{i}-\bar\omega_{i})}{2r}-
    (n+1)\|x'\|_{2}-|p_{i}|.
  \end{align*}
  %
  %
  It is easy to see that for any $ x'\in B_{r}(x) \cap
  \setdef{x'}{\omega_{i}>\bar\omega_{i}^{\operatorname{th}}}$, the first term
  can be arbitrarily large by reducing $r$, while the other two terms
  are bounded; therefore, there exits $r>0$ small enough such
  that~\eqref{eq:auxx} holds.
  By~\eqref{eqn:stability-transient-controller-Lipschitz-4}, this
  implies that $u_{i}(x',p)=0$ for any $ x'\in B_{r}(x)$, and hence
  $u_i$ is Lipschitz in $x$. \qed
\end{pf}

\begin{remark}\longthmtitle{Distributed character and practical
    implementation}\label{rmk:control-realization}
  {\rm The
    controller~\eqref{eqn:stability-transient-controller-Lipschitz-4}
    is distributed since each controlled bus $i \in \CC$, $u_{i}$ only
    utilizes $\omega_{i}$, $p_{i}$, and information of buses it is
    connected to in the power network in order to compute
    $[D^{T}Y_{b}]_{i}\lambda$. This term corresponds to the aggregate
    power flow injected at node~$i$ from its neighboring nodes. In
    turn, this means that, instead of measuring $\lambda_{j}$ and its
    corresponding susceptance for every $i$'s neighboring node $j$, in
    practice, each node can simply measure the signed power flows in
    each neighboring transmission lines of node $i$ and sum it up,
    which is equivalent to $[D^{T}Y_{b}]_{i}\lambda$ as well.}
  \oprocend
\end{remark}

The next result shows that the proposed distributed controller
achieves the objectives identified in
Section~\ref{section:problem-statement} regarding stability,
convergence, and frequency invariance.

\begin{theorem}\longthmtitle{Transient frequency
    control with stability guarantees}\label{thm:decentralized-controller}
  Under condition~\eqref{ineq:sufficient-eq}, let
  $\omega^{\infty}\in(\underline\omega^{\operatorname{th}}_{i},\bar\omega^{\operatorname{th}}_{i})$
  and consider the closed-loop system~\eqref{eqn:dynamics-2} with
  controller~\eqref{eqn:stability-transient-controller-Lipschitz-4}. If
  $\lambda(0)\in \range(D)$ and $(\lambda(0),\omega(0))\in\Phi$ for
  some $\beta>1$, then
  \begin{enumerate}
  \item\label{item:exist-unique} The solution exists and is unique for
    every $t\geqslant 0$;
  \item\label{item:invariance-K} $\lambda(t)\in \range(D)$ and
    $(\lambda(t),\omega(t))\in\Phi$ for any $t\geqslant 0$;
  \item\label{item:convergence-stability-K}
    $(\lambda^{\infty},\omega^{\infty}\ones_{n})$ is stable, and
    $(\lambda(t),\omega(t))\rightarrow(\lambda^{\infty},\omega^{\infty}\ones_{n})$
    as $t\rightarrow \infty$;
    \item\label{item:finite-time-active} The controllers become inactive
    in finite time, i.e., there exists a time $t_{0}>0$ such that
    $u_{i}(x(t),p)=0$ for all $t\geqslant t_{0}$ and all $i\in\CC$.
  \item\label{item:frequency-invariant} For any $i\in\CC$, if
    $\omega_{i}(0)\in[\underline\omega_{i},\bar\omega_{i}]$, then
    $\omega_{i}(t)\in[\underline\omega_{i},\bar\omega_{i}]$ for all
    $t> 0$;
  \item\label{item:frequency-attraction} For any $i\in\CC$, if
    $\omega_{i}(0)\nin[\underline\omega_{i},\bar\omega_{i}]$, then
    $\omega_{i}(t)$ monotonically approaches
    $[\underline\omega_{i},\bar\omega_{i}]$. Furthermore, there exists
    a finite time $t_{1}>0$ such that
    $\omega_{i}(t)\in[\underline\omega_{i},\bar\omega_{i}]$ for all
    $t\geqslant  t_{1}$.
  
  \end{enumerate}
  In addition, if~\ref{item:exist-unique} holds for
  $(\lambda(0),\omega(0))\nin\Phi$,
  then~\ref{item:frequency-invariant} and the monotonic convergence
  in~\ref{item:frequency-attraction} still hold, but with no guarantee
  on the existence of a finite $t_{1}$.
\end{theorem}
\begin{pf}
  It is easy to see
  that~\eqref{eqn:stability-transient-controller-Lipschitz-4}
  guarantees $u_{i}(x,p)\leqslant 0$ if
  $\omega_{i}>\bar\omega_{i}^{\operatorname{th}}$, $u_{i}(x,p)=0$ if
  $\omega_{i}\in(\underline\omega_{i}^{\operatorname{th}},
  \bar\omega_{i}^{\operatorname{th}})$, and $u_{i}(x,p)\geqslant 0$ if
  $\omega_{i}<\underline\omega_{i}^{\operatorname{th}}$.
  Therefore,~\eqref{ineq:stabilize-constraints-2} holds as
  $\omega^{\infty}\in(\underline\omega^{\operatorname{th}}_{i},
  \bar\omega^{\operatorname{th}}_{i})$.  Hence
  \emph{\ref{item:exist-unique}-\ref{item:convergence-stability-K}}
  directly follow from
  Lemma~\ref{lemma:sufficient-stability-convergence}
  (Proposition~\ref{prop:L} justifies the Lipschitzness of the
  controller).
  
  To prove\emph{~\ref{item:finite-time-active}}, we use the
  convergence established
  in\emph{~\ref{item:convergence-stability-K}}. For $\epsilon =
  \min_{i\in\CC}\{\bar\omega_{i}^{\operatorname{th}} -
  \omega^{\infty},\omega^{\infty}-\underline\omega_{i}^{\operatorname{th}}\}$,
  there exists $t_{0}\in\real_{>}$ such that $\|(\lambda(t),\omega(t))
  - (\lambda^{\infty},\omega^{\infty}\ones_{n})\|_{2} < \epsilon$, for
  $ t\geqslant t_{0}$. Therefore, for any $i\in\CC$,
  $|\omega_{i}(t)-\omega^{\infty}| \leqslant
  \|(\lambda(t),\omega(t))-(\lambda^{\infty},\omega^{\infty}\ones_{n})\|_{2}\leqslant
  \min\{\bar\omega_{i}^{\operatorname{th}}-
  \omega^{\infty},\omega^{\infty}-\underline\omega_{i}^{\operatorname{th}}\}$,
  for $ t\geqslant t_{0}$, which implies
  $\underline\omega_{i}^{\operatorname{th}}\leqslant
  \omega_{i}(t)\leqslant\bar\omega_{i}^{\operatorname{th}}$, for $
  t\geqslant t_{0}$.  The result follows now from the
  definition~\eqref{eqn:stability-transient-controller-Lipschitz-4} of
  the controller.  Regarding~\emph{\ref{item:frequency-invariant}},
  the
  controller~\eqref{eqn:stability-transient-controller-Lipschitz-4}
  satisfies~\eqref{ineq:invariance-condition-unsymmetric-3-a} if
  $\bar\omega_{i}^{\operatorname{th}}<\omega_{i}\leqslant
  \bar\omega_{i}$, and
  satisfies~\eqref{ineq:invariance-condition-unsymmetric-3-b} if $
  \underline\omega_{i}\leqslant \omega_{i}<
  \underline\omega_{i}^{\operatorname{th}}$; hence by
  Lemma~\ref{lemma:sufficent-frequecy-invariance} both
  $\bar{\mathcal{C}}_{i}$ and $\underline{\mathcal{C}}_{i}$ are
  invariant.
  
  Proving monotonicity in~\emph{\ref{item:frequency-attraction}} is
  equivalent to showing that $\dot\omega_{i}(t)\leqslant 0$ when
  $\omega_{i}(t)>\bar\omega_{i}$ and $\dot\omega_{i}(t)\geqslant 0$
  when $\omega_{i}(t)<\underline\omega_{i}$. For simplicity we only
  prove the first case. Note that $u_{i}(x,p)\leqslant
  \frac{-\bar\alpha_{i}(\omega_{i} -
    \bar\omega_{i})}{(\omega_{i}-\bar\omega_{i}^{\operatorname{th}})} +
  q_{i}(x,p)$.  Plugging this into~(\ref{eqn:dynamics-2b}) and using
  $\omega_{i}>\bar\omega_{i}$, one has
  \begin{align}\label{ineq:dynamics-bound}
    M_{i}\dot\omega_{i}\leqslant \frac{-\bar\alpha_{i}(\omega_{i} -
      \bar\omega_{i})}{(\omega_{i}-\bar\omega_{i}^{\operatorname{th}})}\leqslant
    0,
  \end{align}
  establishing monotonicity (notice that the inequality holds even if
  the initial condition does not belong to~$\Phi$).  Finally, since
  $\omega^{\infty}\in(\underline\omega^{\operatorname{th}}_{i},
  \bar\omega^{\operatorname{th}}_{i})$ and
  $\omega_{i}(t)\rightarrow\omega^{\infty}$ for every
  $i\in\mathcal{I}$, there exists $t_{1}$ such that $\omega_{i}(t_{1})
  \in
  [\underline\omega^{\operatorname{th}}_{i},\bar\omega^{\operatorname{th}}_{i}]$,
  which, by~\emph{\ref{item:frequency-invariant}}, further implies
  that $\omega(t) \in
  [\underline\omega^{\operatorname{th}}_{i},\bar\omega^{\operatorname{th}}_{i}]$
  for every $t\geqslant t_{1}$.  \qed
  %
  %
\end{pf}

\begin{remark}\longthmtitle{Performance trade-offs via  selection of
    class-$\mathcal{K}$ functions}\label{rmk:linear-class-K}
  {\rm As pointed out in Section~\ref{sec:constraint-freq}, the choice
    of class-$\mathcal{K}$ functions affects the system behavior.  To
    illustrate this, consider the linear choice
    $\bar\alpha_{i}=\underline\alpha_{i}:\real\rightarrow\real,\ s
    \mapsto \gamma_{i}s$, where $\gamma_{i}>0$ is a design parameter.
    A smaller $\gamma_{i}$ leads to more stringent requirements on the
    derivative of the frequency.  This is because $u_{i}(x,p)$ can be
    non-zero only when either of the following happen,
    \begin{align*}
      \frac{-\bar\alpha_{i}(\omega_{i} -
        \bar\omega_{i})}{(\omega_{i}-\bar\omega_{i}^{\operatorname{th}})} +
      q_{i}(x,p)<0\text{ and } \omega_{i}>\bar\omega_{i}^{\operatorname{th}},
      \\
      \frac{\underline\alpha_{i}(\underline\omega_{i} -
        \omega_{i})}{\underline\omega_{i}^{\operatorname{th}}-\omega_{i}} +
      q_{i}(x,p)>0\text{ and }
      \omega_{i}<\underline\omega_{i}^{\operatorname{th}}.
    \end{align*}
    In this first case, the term $\frac{-\bar\alpha_{i}(\omega_{i} -
      \bar\omega_{i})}{(\omega_{i}-\bar\omega_{i}^{\operatorname{th}})} =
    \frac{\gamma_{i}(\bar\omega_{i} -
      \omega_{i})}{\omega_{i}-\bar\omega_{i}^{\operatorname{th}}}>0$ becomes
    smaller as $\gamma_{i}$ decreases, making its addition with
    $q_{i}(x,p)$ more likely to be less than $0$, and resulting in an
    earlier activation of~$u_{i}$.  The second case follows similarly.

    A small $\gamma_{i}$ may also lead to high control magnitude
    because it prescribes a smaller bound on the frequency derivative,
    which in turn may require a larger control effort.
    However, choosing a large $\gamma_{i}$ may cause the controller to
    be highly sensitive to $\omega_{i}$. This is because the absolute
    value of the partial derivative of
    $\frac{-\bar\alpha_{i}(\underline\omega_{i} -
      \omega_{i})}{(\omega_{i}-\bar\omega_{i}^{\operatorname{th}})}$
    (resp. $\frac{\underline\alpha_{i}(\underline\omega_{i} -
      \omega_{i})}{\underline\omega_{i}^{\operatorname{th}}-\omega_{i}}$)
    with respect to $\omega_{i}$ grows proportionally with
    $\gamma_{i}$; consequently, when $u_{i}(x,p)$ is non-zero, its
    sensitivity against $\omega_{i}$ increases as $\gamma_{i}$ grows,
    resulting in low tolerance against slight changes
    in~$\omega_{i}$. In the limit, as $\gamma_{i}\rightarrow \infty$,
    this yields
    \begin{align}\label{eqn:stability-transient-controller-infinite}
      u_{i}^{\infty}(x,p) = 
      \begin{cases}
        \min\{0,q_{i}(x,p)\} & \omega_{i}= \bar\omega_{i},
        \\
        0 & \underline\omega_{i}< \omega_{i}< \bar\omega_{i},
        \\
        \max\{0, q_{i}(x,p)\} & \omega_{i}=\underline\omega_{i},
      \end{cases}    
    \end{align}
    which in general is discontinuous.
    We illustrate in simulation the dependence of the controller on
    the choice of linear class-$\mathcal{K}$ functions in
    Section~\ref{sec:simulations}.}  \oprocend
\end{remark}

\section{Closed-loop performance analysis}\label{sec:performance}
In this section, we characterize additional properties of the
closed-loop system under the proposed distributed controller beyond
stability and frequency invariance. We characterize the attractivity
rate of trajectories for initial conditions outside the safe frequency
region, the boundedness of the control effort prescribed by the
controller along the system trajectories, and its robustness against
measurement and parameter uncertainty.

\subsection{Estimation of the attractivity
  rate}\label{subsection:att-rate}

Here we provide an estimate of the convergence rate to the safe region
(cf.
Theorem~\ref{thm:decentralized-controller}\ref{item:frequency-attraction})
when the frequency of a node is initially outside it.  The next result
identifies a specific trajectory bounding the frequency evolution.

\begin{lemma}\longthmtitle{Upper bound on 
    frequency evolution}\label{lemma:frequency-attractivity}
  With the notation of Theorem~\ref{thm:decentralized-controller},
  assume that for some $i\in\CC$, $\omega_{i}(0)>\bar\omega_{i}$.  Let
  $z_{i}(t)$ be the unique solution of
  \begin{align}\label{eqn:frequency-upper-bound}
    M_{i}\dot z_{i}(t)=
    \frac{-\bar\alpha_{i}(z_{i}(t)-\bar\omega_{i})}{z_{i}(t)-\bar\omega_{i}^{\operatorname{th}}},\
    z_{i}(0)=\omega_{i}(0).
  \end{align}
  Then it holds that $\omega_{i}(t)\leqslant z_{i}(t),$ for any
  $t\geqslant 0$. Furthermore, $z_{i}(t)$ converges to
  $\bar\omega_{i}$ monotonically without reaching it in finite time.
\end{lemma}
\begin{pf}
  It is easy to check that if $z_{i}(0)>\bar\omega_{i}$, then there
  exists a unique solution of~(\ref{eqn:frequency-upper-bound}) for
  every $t\geqslant 0$. Since~\eqref{ineq:dynamics-bound} holds for
  every $i\in\CC$, by the Comparison Lemma~\cite[Lemma~3.4]{HKK:02},
  one has that $\omega_{i}(t)\leqslant z_{i}(t) $ for any $t\geqslant
  0$. On the other hand, one can easily prove via
  Lemma~\ref{lemma:Nagumo} that the set $\left\{ z_{i} \big|
    \bar\omega_{i}-z_{i}\leqslant0 \right\}$ is invariant, which,
  together with the fact that $z_{i}(0)>\bar\omega_{i}$, implies
  $z_{i}(t)\geqslant \omega_{i}$ for every $t\geqslant 0$. By the
  dynamics~\eqref{eqn:frequency-upper-bound}, we deduce $\dot
  z_{i}(t)\leqslant 0$ for every $t\geqslant 0$ and the monotonicity
  follows. Finally, since $z_{i}(t)$ is monotone decreasing and
  lower-bounded, $z_{i}(t)$ is convergent, with limit $\bar\omega_{i}$
  (since $\dot z_{i}(t)<0$ if $z_{i}(t)\neq \bar\omega_{i}$).
  Finally, since the uniqueness of trajectories is guaranteed by the
  Lipschitzness of the dynamics~\eqref{eqn:frequency-upper-bound} and
  $\bar\omega_{i}$ is an equilibrium, it follows that
  $z_{i}(t)>\bar\omega_{i}$ for any $t\geqslant 0$.
  \qed
\end{pf}

A similar statement holds for the case when the initial frequency is
lower than the lower safe bound, but we omit it for brevity.  When
$\bar\alpha_{i}$ is linear, the next result provides an explicit
expression for the bounding trajectory.

\begin{corollary}\longthmtitle{Estimation of frequency convergence rate 
    with linear 
    class-$\mathcal{K}$ function}\label{cor:frequency-upper-bound-linear}
  With the notation of Lemma~\ref{lemma:frequency-attractivity}, if
  $\bar\alpha_{i}(s)=\bar\gamma_{i} s$ with $\bar\gamma_{i}>0$, then
  $z_{i}(t)$ is uniquely determined by
  \begin{align}\label{ineq:upper-bound-z-4}
    z_{i}(t)+(\bar\omega_{i}-\bar\omega_{i}^{\operatorname{th}})
    \ln\left(\frac{z_{i}(t)-\bar\omega_{i}}{\omega_{i}(0)-\bar\omega_{i}}\right)
    = -\bar\gamma_{i}t/M_{i}+\omega_{i}(0).
  \end{align}
  Furthermore, it holds that for any $t\geqslant 0$,
  \begin{align*}
    z_{i}(t)\leqslant \bar\omega_{i} + (\omega_{i}(0)-\bar\omega_{i})
    \exp \Big(\frac{-\bar\gamma_{i} t/M_{i}+
      \omega_{i}(0)-\bar\omega_{i}}{\bar\omega_{i}
      -\bar\omega_{i}^{\operatorname{th}}} \Big) .
  \end{align*}
\end{corollary}
\begin{pf}
  In the case where $\bar\alpha_{i}(s)=\bar\gamma_{i}s$, by separation
  of variables, one has that~\eqref{eqn:frequency-upper-bound} is
  equivalent to
  \begin{align*}
    \frac{z_{i}-\bar\omega_{i}^{\operatorname{th}}}{z_{i}-\bar\omega_{i}}\text{d}z_{i}
    = -\bar\gamma_{i}\text{d}t/M_{i},\ z_{i}(0)=\omega_{i}(0).
  \end{align*}
  Equation~\eqref{ineq:upper-bound-z-4} follows by integrating the
  above differential equation.  Since by
  Lemma~\ref{lemma:frequency-attractivity}
  $z_{i}(t)\geqslant\bar\omega_{i}$ for every $t\geqslant 0$, it holds
  \begin{align*}
    \bar\omega_{i}+(\bar\omega_{i} - \bar\omega_{i}^{\operatorname{th}})
    \ln\left(\frac{z_{i}(t)-\bar\omega_{i}}{\omega_{i}(0)-\bar\omega_{i}}\right)\leqslant
    -\bar\gamma_{i}t/M_{i}+\omega_{i}(0),
  \end{align*}
  concluding the proof. \qed
\end{pf}


\begin{remark}\longthmtitle{Estimation of 
      safe-frequency entry time}\label{rmk:incap-finite-time}
    {\rm Corollary~\ref{cor:frequency-upper-bound-linear} establishes
      the exponential convergence rate of the frequency evolution to
      the safe region, but it does not provide an estimate of the
      finite time of entry $t_1$ stated in
      Theorem~\ref{thm:decentralized-controller}\ref{item:frequency-attraction}. This
      is because the upper-bound signal $z_{i}$ never hits
      $\bar\omega_{i}$ in finite time. This drawback is caused by the
      fact that the existence of $t_{1}$ is justified by (cf.  proof
      of
      Theorem~\ref{thm:decentralized-controller}\ref{item:frequency-attraction})
      the combination of frequency invariance and convergence of the
      closed-loop system, where we do not utilize the latter in
      obtaining the upper-bound signal.  To fix this, one may replace
      $\bar\omega_{i}$ by $\bar\omega_{i}-\epsilon_{i}$
      in~\eqref{eqn:stability-transient-controller-Lipschitz-4} with
      $\epsilon_{i}\in\real_{>}$, and determine $t_{1}$ by solving
      $z(t_{1})=\bar\omega_{i}$ along the
      dynamics~(\ref{eqn:frequency-upper-bound}). Note that, although
      this procedure does not jeopardize any statement in
      Theorem~\ref{thm:decentralized-controller}, it actually puts a
      stricter frequency invariance requirement on the controller.
  } \oprocend
\end{remark}

\subsection{Bounds on controller
  magnitude}\label{subsection:magnitude}
Here, we provide bounds on the amplitude of the proposed
controller~\eqref{eqn:stability-transient-controller-Lipschitz-4}
along the system trajectories for a given constant power injection
profile~$p$.  Our approach to do this is to constrain the allowable
initial conditions by employing the energy function $V$ as a measure
of how far an initial state can be from the equilibrium point.
Formally, let
\begin{align*}
  \hat\Phi(\eta)\triangleq\setdef{(\lambda,\omega)}{\lambda\in
    \Gamma\closure,\ V(\omega,\lambda)\leqslant \eta},
\end{align*}
be the collection of allowable initial states, where $0\leqslant
\eta<c$. The next result bounds the control input as a function
of~$\eta$. 

\begin{lemma}\longthmtitle{Lower bound on control
    effort}\label{lemma:lower-bound}
  For $i\in\CC$, let $g_{i}(\lambda,\omega)\triangleq
  \frac{-\bar\alpha_{i}(\omega_{i} - \bar\omega_{i})}{\omega_{i}
    -\bar\omega_{i}^{\operatorname{th}}} + q_{i}(x,p)$ and
  $d_{i}\triangleq1/2M_{i}(\omega_{i}^{\operatorname{th}} -
  \omega^{\infty})^{2}$. Let $(\lambda^{*},\omega^{*})$ be the optimal
  solution~of
  \begin{subequations}\label{sube:opti-orig}
    \begin{alignat}{2}
      \mathbf{(Q)} \hspace{15mm} &\min_{(\lambda,\omega)} & \quad &
      g_{i}(\lambda,\omega)\notag
      \\
      &\text{s.t.}&\quad &
      (\lambda,\omega)\in\hat\Phi(\eta)\label{opti-orig-1}
      \\
      &&& \lambda\in \range(D) \label{opti-orig-2}
      \\
      &&&
      \omega_{i}>\omega_{i}^{\operatorname{th}} \label{opti-orig-3}
    \end{alignat}
  \end{subequations}
  and define
  \begin{align}
    u_{i}^{\min}(\eta)\triangleq
    \begin{cases}
      0 & \hspace{1.2cm}\text{if $0\leqslant \eta\leqslant d_{i}$,}
      \\
      \min\{0,g_{i}(\lambda^{*},\omega^{*})\} & \hspace{1.2cm}
      \text{if $d_{i}<\eta<c$.}
    \end{cases}\label{case:u-min}
  \end{align}
  %
  %
  Then, for any $(\lambda(0),\omega(0))\in\hat\Phi(\eta)$ with
  $\lambda(0)\in \range(D)$,
  \begin{align}\label{ineq:lower-bound}
    u_{i}(x(t),p)\geqslant u_{i}^{\min}(\eta),
  \end{align}
  for any $t\geqslant 0$, and there exists initial states such that
  equality holds at some $t\geqslant 0$.
\end{lemma}
\begin{pf}
  Note that by Theorem~\ref{thm:decentralized-controller} with
  $\beta=c/\eta>1$, one has $(\lambda(t),\omega(t))\in\hat\Phi(\eta)$
  and $\lambda(t)\in \range(D)$ for every $t> 0$, provided they hold
  at~$t=0$. Therefore, to show~\eqref{ineq:lower-bound} for every
  $t\geqslant 0$, it suffices to show it holds for~$t=0$.  If
  $0\leqslant \eta\leqslant d$, then
  $1/2M_{i}(\omega_{i}(0)-\omega^{\infty})^{2}\leqslant
  V(\omega(0),\lambda(0))\leqslant d_{i}=
  1/2(M_{i}(\omega_{i}^{\operatorname{th}}-\omega^{\infty})^{2}$,
  which implies $\omega_{i}(0)\leqslant
  \omega_{i}^{\operatorname{th}}$; therefore, $u_{i}(x(0),p)\geqslant
  0$ follows
  by~\eqref{eqn:stability-transient-controller-Lipschitz-4}. Also,
  $u_{i}(x(0),p)$ can be $0$ in the case when, say,
  $x(0)=(\lambda^{\infty},\omega^{\infty})$. In the other case, if
  $d_{i}<\eta<c$, then $u_{i}(x(0),p)$ is lower bounded by the optimal
  value of
  \begin{alignat}{2}\label{sube:u-bound}
    \mathbf{(\hat{Q})} \hspace{15mm} &\min_{\lambda,\omega} & \quad &
    u_{i}(x,p)\notag
    \\
    &\text{s.t.}&\quad &~\eqref{opti-orig-1}\text{
      and}~\eqref{opti-orig-2}.
  \end{alignat}
  Denote this optimal value by $v_{i}(\eta)$.  Also, the value of
  $u_{i}(x(0),p)$ can be exactly $v_{i}(\eta)$, e.g., in the case when
  $x(0)$ is the optimal solution of $(\hat Q)$. Note that $v_{i}(\eta)
  \leqslant 0$ as $(\lambda^{\infty},\omega^{\infty})$
  satisfies~\eqref{sube:u-bound} and
  $u_{i}((\lambda^{\infty},\omega^{\infty}),p)=0$.  Since it holds
  that a) $u_{i}(x,p)\geqslant 0$ for any $\omega_{i}\leqslant
  \omega_{i}^{\operatorname{th}}$, and b) $u_{i}(x,p)\leqslant 0$ for
  any $\omega_{i}\geqslant \omega_{i}^{\operatorname{th}}$, one can,
  without changing the optimal value, replace $u_{i}(x,p)$ by
  $\min\{0,g_{i}(\lambda,\omega)\}$ in $(\hat Q)$, and meanwhile add
  an additional constraint~\eqref{opti-orig-3}.  With a simple
  reasoning effort, one can show that for this new optimization
  problem, the optimal value is exactly
  $\min\{0,g_{i}(\lambda^{*},\omega^{*})\}$. \qed
\end{pf}

Note that the control amplitude lower bound $u_{i}^{\min}(\eta)$
depends nonlinearly on the power injection~$p$. This is because,
although the objective function in the optimization problem $(Q)$,
linearly depends on $p$, the optimal value does depend nonlinearly on
$p$ through the constraint~\eqref{opti-orig-1}. This is due to the
fact that the equilibrium
$(\lambda^{\infty},\omega^{\infty}\ones_{n})$ depends on $p$ through
the transcendental equation~\eqref{eqn:lambda-solution}.


  
A similar result can be stated regarding an upper bound of the
controller magnitude, but we omit it for brevity.  The problem $(Q)$
is non-convex due to the non-convexity of the objective function.  We
next show that its optimal value equals that of another optimization
problem with convex objective function and non-convex feasible set.
Define the function $h_{i}:\real^{m+n}\times\real\rightarrow\real^{},\
(z,\omega)\rightarrow h_{i}(z,\omega)$ exactly the same as $g_{i}$ but
replacing $\sin\lambda_{i}$ by $z_{i}$ in the definition of~$q_i$. In
this way, $h_{i}(\sin\lambda,\omega) = g_{i}(\lambda,\omega)$. Let
$\mathcal{D}_{i}^{+}\triangleq\left\{ j \big| [D^{T}Y_{b}]_{ij}> 0
\right\}$ and $\mathcal{D}_{i}^{-}\triangleq\left\{ j \big|
  [D^{T}Y_{b}]_{ij}<0 \right\}$. Consider the optimization
\begin{subequations}\label{sube:opti:lossless-relax}
  \begin{alignat}{2}
    \mathbf{(R)} \hspace{15mm}
    &\min_{(z,\lambda,\omega)} & \quad & h_{i}(z,\omega)\notag
    \\
    &\text{s.t.}&\quad &\sin\lambda_{j}\leqslant z_{j},\ \forall
    j\in\mathcal{D}^{+}_{i},\label{ineq:positive-relax}
    \\
    &&& \sin\lambda_{j}\geqslant z_{j},\ \forall
    j\in\mathcal{D}^{-}_{i}, \label{ineq:negative-relax}
    \\
    &&& \eqref{opti-orig-1}\text{ to}~\eqref{opti-orig-3},
  \end{alignat}
\end{subequations}
We claim that the optimal value of this problem is the same as that of
$(Q)$.  
The claim holds if every optimal solution of $(R)$, denoted by
$(z^{\sharp},\lambda^{\sharp},\omega^{\sharp})$,
satisfies~\eqref{ineq:positive-relax} and~\eqref{ineq:negative-relax}
with equality signs. This has to be the case since, for instance, if
$\sin\lambda_{k}^{\sharp}<z_{k}^{\sharp}$ for some
$k\in\mathcal{D}^{+}_{i}$, then
$(z^{\sharp},\lambda^{\sharp},\omega^{\sharp})$ can no more be an
optimal solution, since $(\hat
z^{\sharp},\lambda^{\sharp},\omega^{\sharp})$, where $\hat z^{\sharp}$
differs from $z^{\sharp}$ only in its $k$th component, $\hat
z^{\sharp}_{k} = \sin\lambda_{k}^{\sharp}$, has $h_{i}(\hat
z^{\sharp},\omega^{\sharp})<h_{i}(z^{\sharp},\omega^{\sharp})$,
violating optimality.

Our next step is to convexify $(R)$. Here we assume that
$\omega_{i}\mapsto \frac{-\bar\alpha_{i}
  (\omega_{i}-\bar\omega_{i})}{\omega_{i} -
  \bar\omega_{i}^{\operatorname{th}}}$ is convex in $\omega_{i}$ in
the region $\omega_{i}>\omega_{i}^{\operatorname{th}}$, which suffices
to guarantee the convexity of $(z,\omega)\mapsto h_{i}(z,\omega)$ in
$(z,\omega)$ under constraint~\eqref{sube:opti:lossless-relax} (this
convexity assumption holds if, for instance, $\bar\alpha_{i}$ is a
linear function).
To handle the non-convexity of the
constraints~\eqref{ineq:positive-relax}
and~\eqref{ineq:negative-relax}, in the following two results, we
separately provide inner and outer approximations,
%
%
leading to upper and lower approximations of the optimal value
of~$(R)$, and equivalently~$(Q)$.

\begin{figure}[htb]
  \centering%
  \subfigure[\label{fig:sin-inner}]{\includegraphics[width=.49\linewidth]{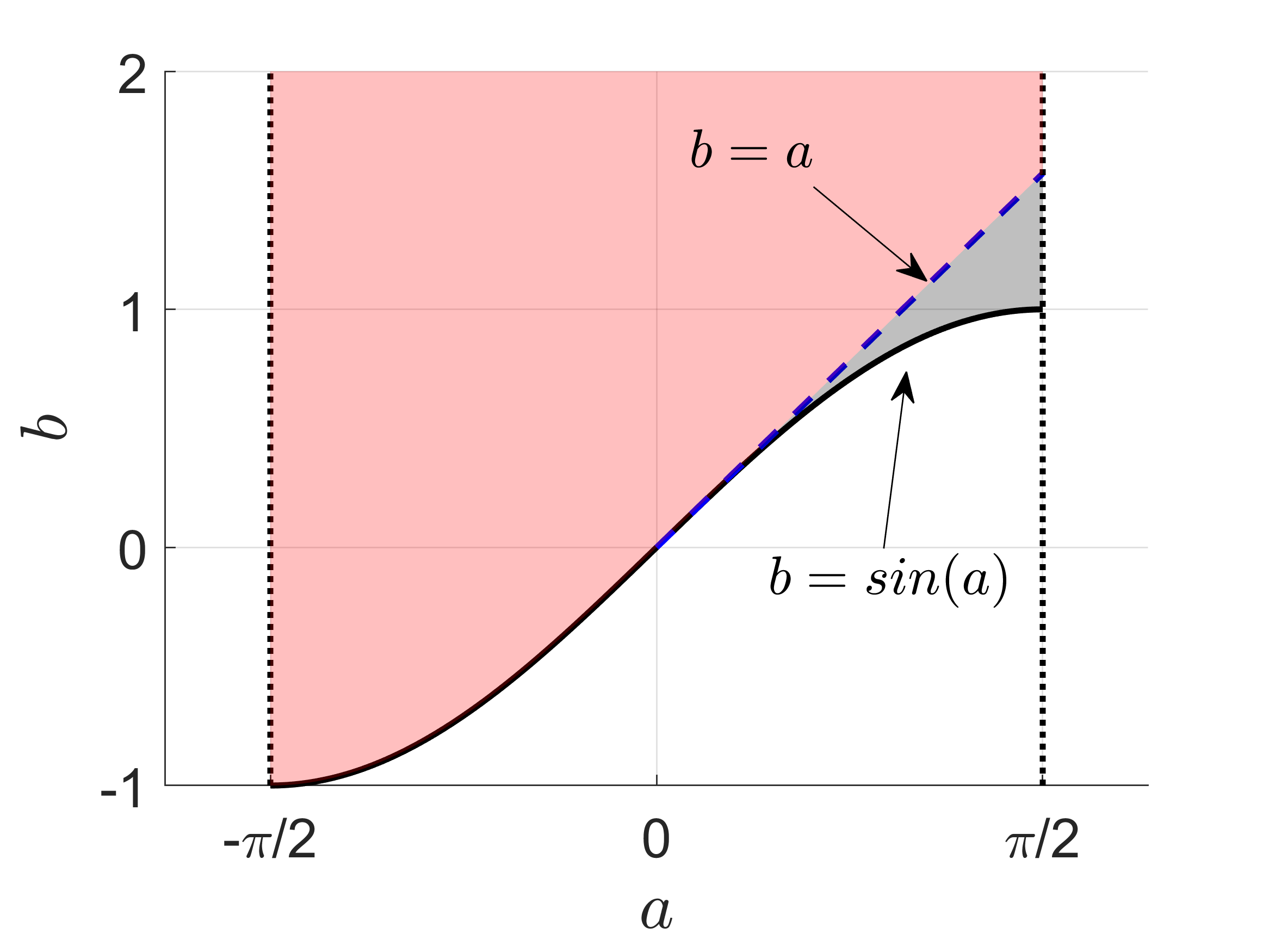}}
  \subfigure[\label{fig:sin-outer}]{\includegraphics[width=.49\linewidth]{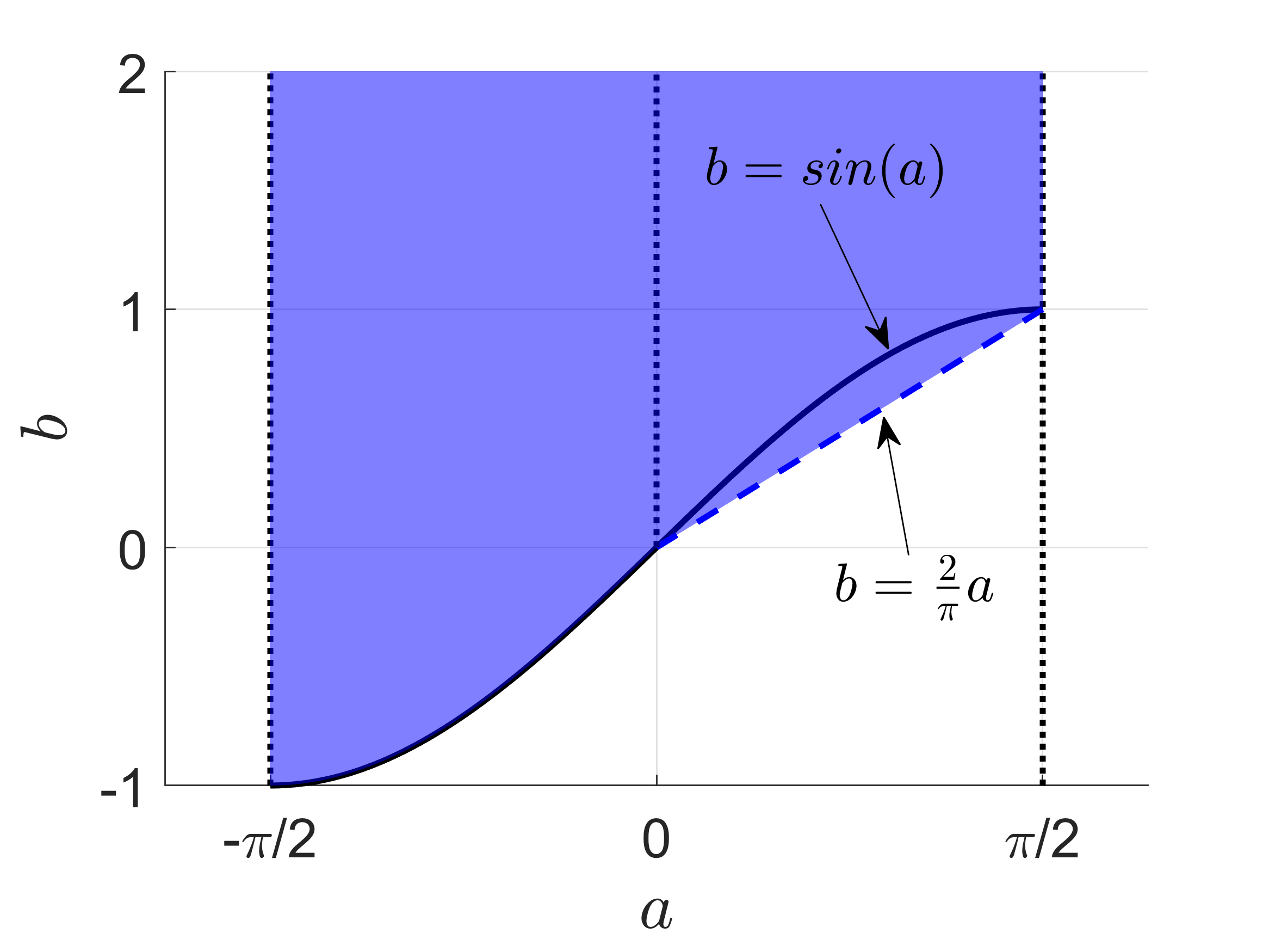}}
  \caption{Tightening and relaxation of a sinusoidal non-convex
    constraint. In plot~\subref{fig:sin-inner}, within $|a|<\pi/2$, by
    ignoring the gray region delimited by $b=a$, $b=\sin (a)$ and
    $a=\pi/2$, the non-convex set characterized by $\sin(a)\leqslant
    b$ appearing in~\eqref{ineq:positive-relax} contains the red
    convex subset~$\mathcal{H}^{+}$. On the other hand, in
    plot~\subref{fig:sin-outer}, this non-convex set is contained in
    the blue region. Each of the blue regions separated by the dotted
    line at $a=0$ are convex.}\label{fig:sin-appr}
\end{figure}

\begin{lemma}\longthmtitle{Upper bound of optimal
    value}\label{lemma:upper-optimal}
  Define $\mathcal{H}^{+} \triangleq \{ (a,b)\big|\ |a|<\pi/2,\ \sin
  a\leqslant b\text{ if }a\in[-\pi/2,0),\text{ and } a\leqslant
  b\text{ if } a\in[0,\pi/2] \}$, and $\mathcal{H}^{-}\triangleq \{
  (a,b)\big|\ |a|<\pi/2,\ a\geqslant b\text{ if }a\in[-\pi/2,0),\text{
    and}$ $ \sin a\geqslant b\text{ if } a\in[0,\pi/2] \}$.  Consider
  the convex optimization problem
  \begin{subequations}\label{sube:opti:lossless-relax-2}
    \begin{alignat}{2}
      \mathbf{(\bar R)} \hspace{15mm} &\min_{(z,\lambda,\omega)} &
      \quad & h_{i}(z,\omega)\notag
      \\
      &\text{s.t.}&\quad &(\lambda_{j}, z_{j})\in\mathcal{H}^{+},\
      \forall j\in\mathcal{D}^{+}_{i},\label{ineq:positive-relax-2}
      \\
      &&& (\lambda_{j}, z_{j})\in\mathcal{H}^{-},\ \forall
      j\in\mathcal{D}^{-}_{i}, \label{ineq:negative-relax-2}
      \\
      &&& \eqref{opti-orig-1}\text{ to}~\eqref{opti-orig-3},
    \end{alignat}
  \end{subequations}
  and denote its optimal solution by
  $(z^{o},\lambda^{o},\omega^{o})$. Then it holds that
  $h_{i}(z^{o},\omega^{o})\geqslant
  g_{i}(\lambda^{o},\omega^{o})\geqslant
  g_{i}(\lambda^{*},\omega^{*})$.
\end{lemma}
\begin{pf}
  The second inequality holds since $(\lambda^{o},\omega^{o})$
  satisfies~\eqref{opti-orig-1}\text{ to}~\eqref{opti-orig-3}, making
  it a feasible point for~$(Q)$. To show the first inequality, one can
  easily check that for any $j\in\mathcal{D}^{+}_{i}$, if
  $(\lambda_{j},z_{j})\in\mathcal{H}^{+}$, then $\sin
  \lambda_{j}\leqslant z_{j}$ (cf. Figure~\ref{fig:sin-inner}).
  Therefore,~\eqref{ineq:positive-relax-2} is stricter
  than~\eqref{ineq:positive-relax}.
  Similarly,~\eqref{ineq:negative-relax-2} is stricter
  than~\eqref{ineq:negative-relax}. Therefore,
  $[D^{T}Y_{b}]_{ij}z^{o}_{j}\geqslant
  [D^{T}Y_{b}]_{ij}\sin\lambda_{j}^{o}$ holds for any
  $j\in[1,m]_{\naturals}$, completing the proof since
  $h_{i}(z^{o},\omega^{o})\geqslant
  h_{i}(\sin\lambda^{o},\omega^{o})=g_{i}(\lambda^{o},\omega^{o})$. \qed 
\end{pf}

\begin{lemma}\longthmtitle{Lower bound of optimal
    value} \label{lemma:lower-optimal}
  Define $\mathcal{M}^{+}_{0}\triangleq \{ (a,b)\big|\
  -\pi/2<a\leqslant 0,\ \sin a\leqslant b\}$,
  $\mathcal{M}^{+}_{1}\triangleq \{ (a,b)\big|\ 0\leqslant a\leqslant
  \pi/2,\ 2a/\pi \leqslant b\}$, $\mathcal{M}^{-}_{0}\triangleq\{
  (a,b)\big|\ -\pi/2<a\leqslant 0,\ 2a/\pi \geqslant b\}$, and
  $\mathcal{M}^{-}_{1}\triangleq\{ (a,b)\big|\ 0\leqslant a\leqslant
  \pi/2,\ \sin a \leqslant b\}$.  Consider the convex optimization
  problem for $\mu \triangleq
  \{\mu_{j}\}_{j\in\mathcal{D}^{+}_{i}\bigcup\mathcal{D}^{-}_{i}}$,
  with $\mu_{j}\in\{0,1\}$,
  \begin{subequations}\label{sube:opti:relax-3}
    \begin{alignat}{2}
      \mathbf{(\underline R^{\mu})} \hspace{15mm}
      &\min_{(z,\lambda,\omega)} & \quad & h_{i}(z,\omega)\notag
      \\
      &\text{s.t.}&\quad &(\lambda_{j},
      z_{j})\in\mathcal{M}^{+}_{\mu_{j}},\ \forall
      j\in\mathcal{D}^{+}_{i},\label{ineq:positive-relax-3}
      \\
      &&& (\lambda_{j}, z_{j})\in\mathcal{M}^{-}_{\mu_{j}},\ \forall
      j\in\mathcal{D}^{-}_{i}, \label{ineq:negative-relax-3}
      \\
      &&& \eqref{opti-orig-1}\text{ to}~\eqref{opti-orig-3},
    \end{alignat}
  \end{subequations}
  and denote its optimal solution by $(\underline
  z^{\mu},\underline\lambda^{\mu},\underline\omega^{\mu})$. Let
  $\mu^{*}\triangleq\arg\min_{\mu}h_{i}(\underline
  z^{\mu},\underline\omega^{\mu})$, then $h_{i}(\underline
  z^{\mu^{*}},\underline\omega^{\mu^{*}})\leqslant
  g_{i}(\lambda^{*},\omega^{*})$.
\end{lemma}
\begin{pf}
  Define 
  \begin{subequations}\label{opti:relax-4}
    \begin{alignat}{2}
      \mathbf{(\underline R)} \hspace{15mm} &\min_{(z,\lambda,\omega)}
      & \quad & h_{i}(z,\omega)\notag
      \\
      &\text{s.t.}&\quad &(\lambda_{j}, z_{j})\in\mathcal{M}^{+}_{0}
      \cup \mathcal{M}^{+}_{1},\ \forall
      j\in\mathcal{D}^{+}_{i},\label{ineq:positive-relax-4}
      \\
      &&& (\lambda_{j}, z_{j})\in\mathcal{M}^{-}_{0} \cup
      \mathcal{M}^{-}_{1},\ \forall
      j\in\mathcal{D}^{-}_{i}, \label{ineq:negative-relax-4}
      \\
      &&& \eqref{opti-orig-1}\text{ to}~\eqref{opti-orig-3}.
    \end{alignat}
  \end{subequations}
  One can easily see
  that~\eqref{ineq:positive-relax}-\eqref{ineq:negative-relax} is
  stricter
  than~\eqref{ineq:positive-relax-4}-\eqref{ineq:negative-relax-4}
  (cf.  Figure~\ref{fig:sin-outer}). Hence the optimal value of
  $(\underline R)$ lower bounds
  $g_{i}(\lambda^{*},\omega^{*})$. Notice
  that~\eqref{ineq:positive-relax-3}-\eqref{ineq:negative-relax-3}
  simply
  splits~\eqref{ineq:positive-relax-4}-\eqref{ineq:negative-relax-4}
  into convex regions, and hence $(\underline
  z^{\mu^{*}},\underline\lambda^{\mu^{*}},\underline\omega^{\mu^{*}})$
  is also the optimal solution of $(\underline R^{{\mu}})$. \qed
\end{pf}

Together, Lemmas~\ref{lemma:upper-optimal}
and~\ref{lemma:lower-optimal} provide us with efficient ways of
approximating the value of the bound on the control
effort~$u_{i}^{\min}(\eta)$.

\subsection{Robustness to measurement and parameter uncertainty} 

Here we study the controller performance under measurement and
parameter uncertainty. This is motivated by scenarios where the state
or the power injection may not be precisely measured, or scenarios
where some system parameters, like the damping coefficient, are only
approximately known.  Formally, we let $\hat
x=(\hat\lambda,\hat\omega)$, $\hat p$, and $\hat E$ be the measured or
estimated state, power injection, and damping parameters,
respectively. For every $i\in\CC$, we introduce the error variables
\begin{align*}
  \epsilon^{\omega}_{i}&\triangleq
  \hat\omega_{i}-\omega_{i},\
  &\epsilon^{\lambda}_{i}\triangleq
  [D^{T}Y_{b}]_{i}\hat\lambda-[D^{T}Y_{b}]_{i}\lambda,
  \\
  \epsilon^{p}_{i}&\triangleq \hat p_{i}-p_{i},\
  &\epsilon^{E}_{i}\triangleq \hat E_{i}-E_{i}.\hspace{1.94cm}
\end{align*}
We make the following assumption regarding the error.

\begin{assumption}\longthmtitle{Bounded
    uncertainties}\label{assumption:bounded-uncertain}
  For each $i\in\CC$,
  \begin{enumerate}
  \item\label{item:uncertain-bound} the uncertainties are piece-wise
    continuous and can be bounded by
    $|\epsilon^{\omega}_{i}(t)|\leqslant \bar\epsilon^{\omega}_{i}$,
    $|\epsilon^{\lambda}_{i}(t)|\leqslant \bar\epsilon^{\lambda}_{i}$,
    $|\epsilon^{p}_{i}(t)|\leqslant \bar\epsilon^{p}_{i}$, and
    $|\epsilon^{E}_{i}(t)|\leqslant \bar\epsilon^{E}_{i}$ for all
    $t\geqslant 0$;
  \item\label{item:robust-frequency-bound} $\omega^{\infty} \in
    (\underline\omega_{i}^{\operatorname{th}} +
    \bar\epsilon^{\omega}_{i},\bar\omega_{i}^{\operatorname{th}}-\bar\epsilon^{\omega}_{i})$;
  \item\label{item:uncertain-freqeuncy-bound} $\bar\epsilon^{\omega}_{i}<
    \min\{\bar\omega_{i}-\bar\omega_{i}^{\operatorname{th}},
    \underline\omega_{i}^{\operatorname{th}}-\underline\omega_{i}\}$.
  \end{enumerate}
\end{assumption}

Condition~\emph{\ref{item:uncertain-bound}} provides uniform bounds on
the uncertainties;~\emph{\ref{item:robust-frequency-bound}} ensures
that, even with uncertainty, the control input is identically 0 around
the equilibrium;~\emph{\ref{item:uncertain-freqeuncy-bound}}
guarantees that the control input is always non-singular.

For convenience, we use~$\hat u_{i}(\hat x,\hat p(t))$ to refer to the
controller with the same functional expression
as~\eqref{eqn:stability-transient-controller-Lipschitz-4} but
implemented with approximate parameter values and evaluated at the
inaccurate state $\hat x$ and power injection $\hat p(t)$. Notice that
$\hat p(t)$ can be time-varying.  The next result shows that $\hat
u_{i}$ still stabilizes the power network and enforces the
satisfaction of a relaxed frequency invariance condition. For
simplicity, we restrict our attention to linear class-$\mathcal{K}$
functions in the controller design.


\begin{proposition}\longthmtitle{Robust stability and frequency
    invariance under uncertainty}\label{prop:robust-uncertainty}
  Under condition~\eqref{ineq:sufficient-eq} and
  Assumption~\ref{assumption:bounded-uncertain}, consider the
  evolution of the system~\eqref{eqn:dynamics-2} with the controller
  $\hat u_{i}$ for each $i\in\CC$. Then the following results hold
  provided $\lambda(0) \in \range(D)$ and
  $(\lambda(0),\omega(0))\in\Phi$ for some $\beta>1$:
  \begin{enumerate}
  \item\label{item:robust-sol-existence} The solution exists and is
    unique for every $t\geqslant 0$.
  \item\label{item:robust-invariance-K} $\lambda(t)\in \range(D)$ and
    $(\lambda(t),\omega(t))\in\Phi$ for any $t\geqslant 0$;
  \item\label{item:robust-stability} $(\lambda^{\infty},\
    \omega^{\infty}\ones_{n})$ is stable, and $\left(\lambda(t),\
      \omega(t)\right)$ converges to $(\lambda^{\infty},\
    \omega^{\infty}\ones_{n})$;
  \item\label{item:robust-finite-time} There exists a finite time
    $t_{2}$ such that $\hat u_{i}(\hat x(t),\hat p(t))=0$ for every
    $t\geqslant t_{2}$ and every $i\in\CC$.
  \item\label{item:robust-invariance} Suppose $\bar\alpha_{i}(s) =
    \underline\alpha_{i}(s)=\gamma_{i}s$ for every $i\in\CC$.  Then,
    if there exists $\Delta>0$ such that satisfy
    \begin{subequations}\label{sube:ineq:robust-invariance}
      \begin{align}
        \hspace{-1.2cm} \frac{-\gamma_{i}(\bar\epsilon^{\omega}_{i} +
          \Delta)}{\bar\omega_{i}-\bar\omega_{i}^{\operatorname{th}} +
          \Delta+\bar\epsilon^{\omega}_{i}}+\bar\epsilon^{E}_{i}(\Delta+\bar\omega_{i})
        + \hat
        E_{i}\bar\epsilon^{\omega}_{i}+\bar\epsilon^{\lambda}_{i} +
        \bar\epsilon^{p}_{i}\leqslant
        0,\label{sube:ineq:robust-invariance-a}
        \\
        \hspace{-1.2cm} \frac{-\gamma_{i}(\bar\epsilon^{\omega}_{i} +
          \Delta)}{\underline\omega_{i}^{\operatorname{th}}-\underline\omega_{i}
          + \Delta + \bar\epsilon^{\omega}_{i}} +
        \bar\epsilon^{E}_{i}(\Delta-\underline\omega_{i})+\hat
        E_{i}\bar\epsilon^{\omega}_{i} +
        \bar\epsilon^{\lambda}_{i}+\bar\epsilon^{p}_{i}\leqslant
        0,\label{sube:ineq:robust-invariance-b}
      \end{align}
      then $\omega_{i}(t) \in [\underline\omega_{i}-\Delta,
      \bar\omega_{i}+\Delta]$ for all $t> 0$, provided $\omega_{i}(0)
      \in [\underline\omega_{i}-\Delta,\bar\omega_{i}+\Delta]$, and,
      if $\omega_{i}(0)\nin
      [\underline\omega_{i}-\Delta,\bar\omega_{i}+\Delta]$, then there
      exists a finite time $t_{3}$ such that $\omega_{i}(t) \in
      [\underline\omega_{i}-\Delta, \bar\omega_{i}+\Delta]$ for all
      $t\geqslant t_{3}$.
    \end{subequations}
  \end{enumerate}
\end{proposition}
\begin{pf}  
  The proofs
  of~\emph{\ref{item:exist-unique}-\ref{item:robust-stability}} follow
  similar arguments as the proofs of
  Theorem~\emph{\ref{item:exist-unique}-\ref{item:convergence-stability-K}}.
  For stability,
  one can show that $\frac{d}{dt}V(\omega(t),\lambda(t)) =
  -\tilde\omega^{T}(t)E\tilde\omega(t)
  +\sum_{i\in\CC}\tilde\omega_{i}(t)\hat u_{i}(\hat x(t),\hat p(t))$.
  By Assumption~\ref{assumption:bounded-uncertain} and the definition
  of $\hat u_{i}$, it holds that
  $\sum_{i\in\CC}\tilde\omega_{i}(t)\hat u_{i}(\hat x(t),\hat
  p(t))\leqslant 0$, implying
  $\frac{d}{dt}V(\lambda(t),\omega(t))\leqslant 0$. The convergence
  follows by LaSalle Invariance Principle and noticing that $\hat
  u_{i}(\hat x,\hat p (t))$ is identically 0 so long as $\omega_{i}\in
  [\underline\omega_{i}^{\operatorname{th}}+\bar\epsilon^{\omega}_{i},\bar\omega_{i}^{\operatorname{th}}-\bar\epsilon^{\omega}_{i}]$,
  which, together with the convergence, implies that $\hat u_{i}(\hat
  x(t),\hat p (t))$ is 0 after a finite time.
%
  For~\emph{\ref{item:robust-invariance}}, to prove the invariance of
  $[\underline\omega_{i}-\Delta,\bar\omega_{i}+\Delta]$, by
  Lemma~\ref{lemma:frequency-invariance}, we only need to show that
  \begin{subequations}
    \begin{align}
      \hat u_{i}(\hat x,\hat p(t))-q_{i}(x,t)\leqslant 0,\
      \text{if }
      \omega_{i}=\bar\omega_{i}+\Delta,\label{ineq:robust-invariance-1a}
      \\
      -\hat u_{i}(\hat x,\hat p(t))+q_{i}(x,t)\leqslant 0,\
      \text{if }\omega_{i}=\underline
      \omega_{i}-\Delta.\label{ineq:robust-invariance-1b}
    \end{align}\label{ineq:robust-invariance}
  \end{subequations}
  For simplicity, we only show
  that~\eqref{sube:ineq:robust-invariance-a}
  implies~\eqref{ineq:robust-invariance-1a} (the fact
  that~\eqref{sube:ineq:robust-invariance-b}
  implies~\eqref{ineq:robust-invariance-1b} follows similarly). Notice
  that if $\omega_{i}=\bar\omega_{i}+\Delta$, then $\hat u_{i}(\hat
  x,\hat p(t))-q_{i}(x,t)$ equals
  \begin{align}
    \frac{-\gamma_{i}(\Delta+\epsilon^{\omega}_{i})}{\bar\omega_{i} -
      \bar\omega_{i}^{\operatorname{th}} +
      \Delta+\epsilon^{\omega}_{i}}+\epsilon^{E}_{i}(\bar\omega_{i}+\Delta)
    + \hat E_{i}\epsilon^{\omega}_{i} + \epsilon^{\lambda}_{i} +
    \epsilon^{p}_{i},\label{exp:omega-dot}
  \end{align}
  which, by Assumption~\ref{assumption:bounded-uncertain}, is smaller
  than or equal to the left-hand side
  of~\eqref{sube:ineq:robust-invariance-a} by letting the
  uncertainties take their individual bounds;
  hence~\eqref{ineq:robust-invariance-1a} holds. Finally, the
  existence of $t_{3}$ follows a similar proof in
  Theorem\emph{~\ref{item:frequency-attraction}}. \qed
\end{pf}

One should look at~\eqref{sube:ineq:robust-invariance} as a condition
that, independently of the specific realization of the uncertainty,
guarantees that the invariance of the frequency interval is ensured.


\begin{figure}[tb!]
  \centering%
  \includegraphics[width=1.1\linewidth]{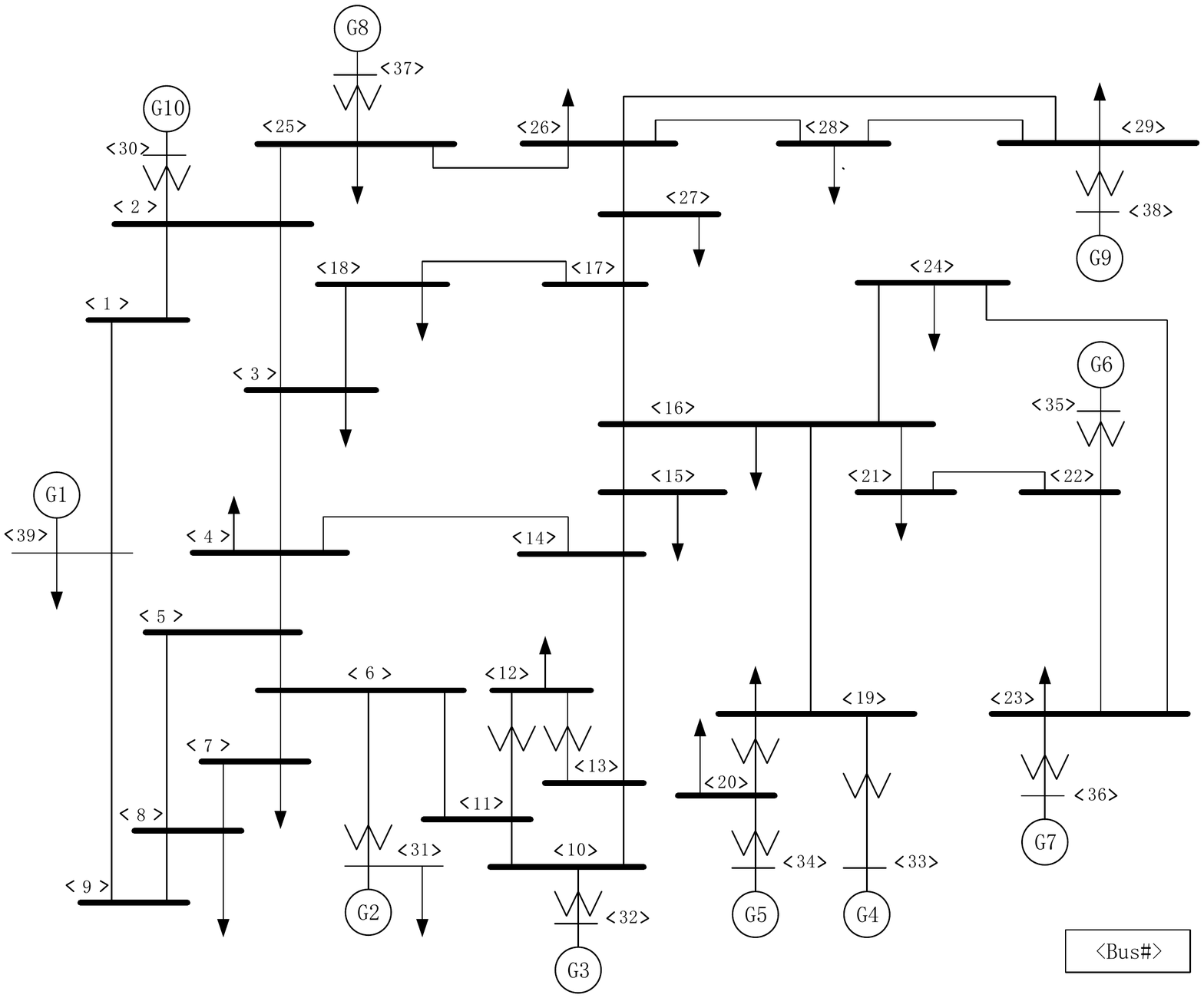}
  \caption{IEEE 39-bus power network.}\label{fig:IEEE39bus}
\end{figure}

\section{Simulations}\label{sec:simulations}

We illustrate the performance of our control design in the IEEE 39-bus
power network displayed in Figure~\ref{fig:IEEE39bus}.  The network
consists of 46 transmission lines and 10 generators, serving a load of
approximately 6GW.  We take the values of susceptance $b_{ij}$ and
rotational inertia~$M_{i}$ for generator nodes from the Power System
Toolbox~\citep{KWC-JC-GR:09}.  We use this toolbox to assign the
initial power injection $p_{i}(0)$ for every bus (although the
analytical results hold for constant power injections, in simulation
we have also tested the more general time-varying case). We assign all
non-generator buses a uniform small inertia $M_{i}=0.1$.  The damping
parameter is $E_{i}=1$ for all buses.  The initial state $(\lambda
(0),\omega(0))$ is chosen to be the unique equilibrium with respect to
the initial power injection.  We implement the distributed controller
in~\eqref{eqn:stability-transient-controller-Lipschitz-4} in the
generators with indices $\CC=\{30,31,32\}$ to tune their transient
frequency behavior. The controller parameters are as follows: for
every $i \in \CC$, we let $\bar\alpha_{i}(s) =
\underline\alpha_{i}(s)=\gamma_{i}s$, with $\gamma_{i}=2$,
$\bar\omega_{i} = -\underline\omega_{i} = 0.2$Hz and
$\bar\omega_{i}^{\operatorname{th}} =
-\underline\omega_{i}^{\operatorname{th}} = 0.1$Hz.  The nominal
frequency is 60Hz, and hence the safe frequency region is
$[59.8\text{Hz},\ 60.2\text{Hz}]$.

\begin{figure*}[tbh!]
  \centering
  \includegraphics[width=1\linewidth,height=0.25\linewidth]{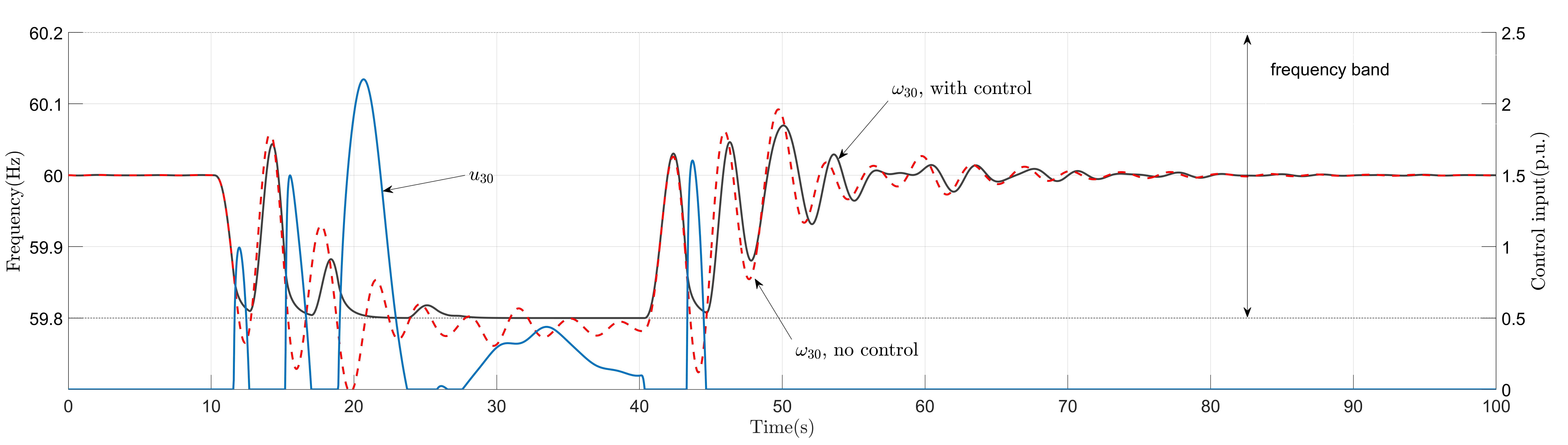}
  \caption{Frequency and control input trajectories at node 30
    corresponding to the power supply loss of generator G9 during
    [10,40]s. The frequency trajectory without transient controller
    goes beyond the safe bounds during the contingency, while this is
    avoided with the proposed controller. Notice that the latter xonly
    takes effect when the frequency is close to the safe
    bound.}\label{fig:generator-loss}
\end{figure*}

\begin{figure*}[tbh!]
  \centering
  \subfigure[\label{frequency-response-no-control-generator}]{\includegraphics[width=.23\linewidth]{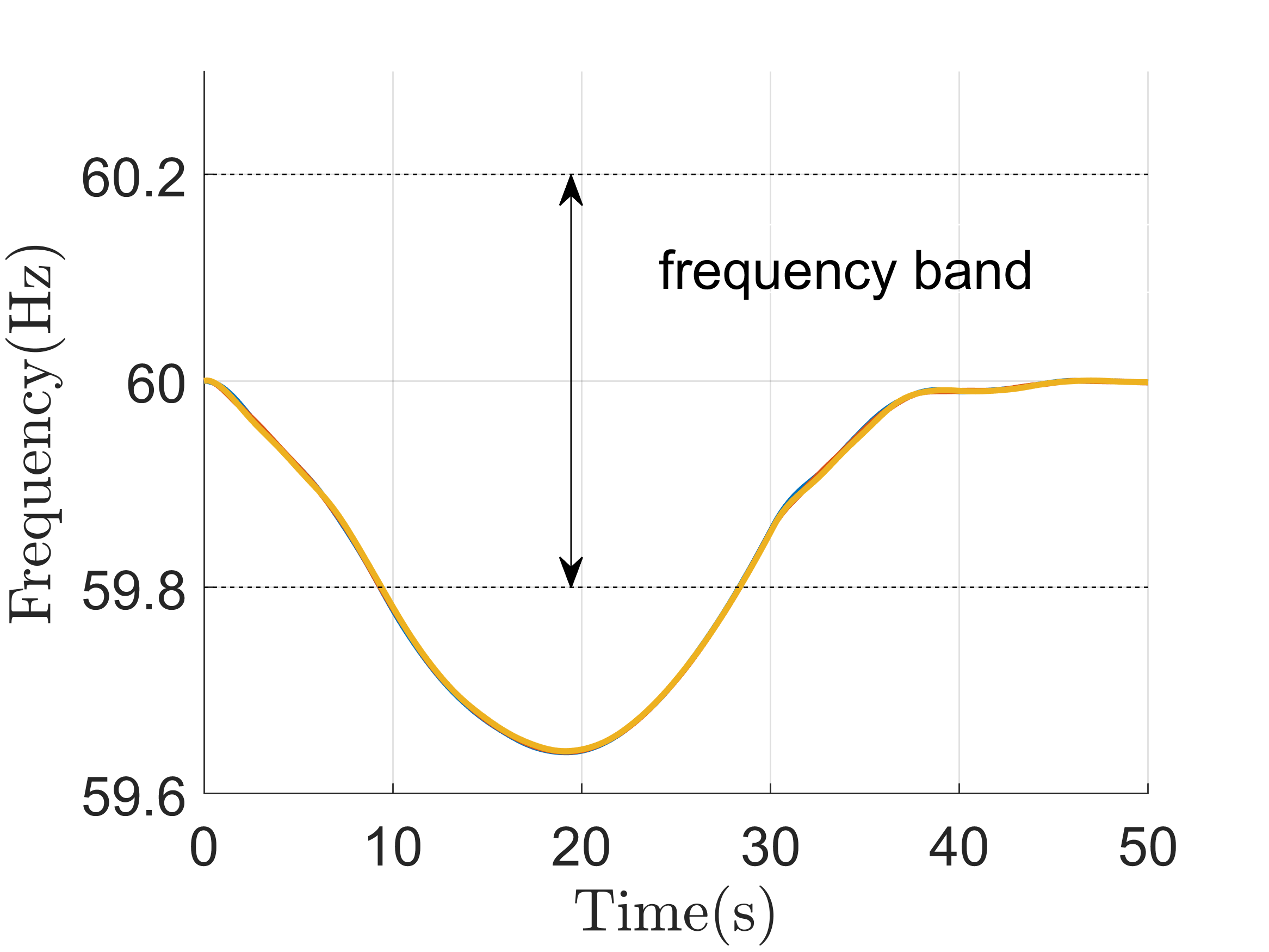}}
  \subfigure[\label{frequency-response-with-control-generator}]{\includegraphics[width=.23\linewidth]{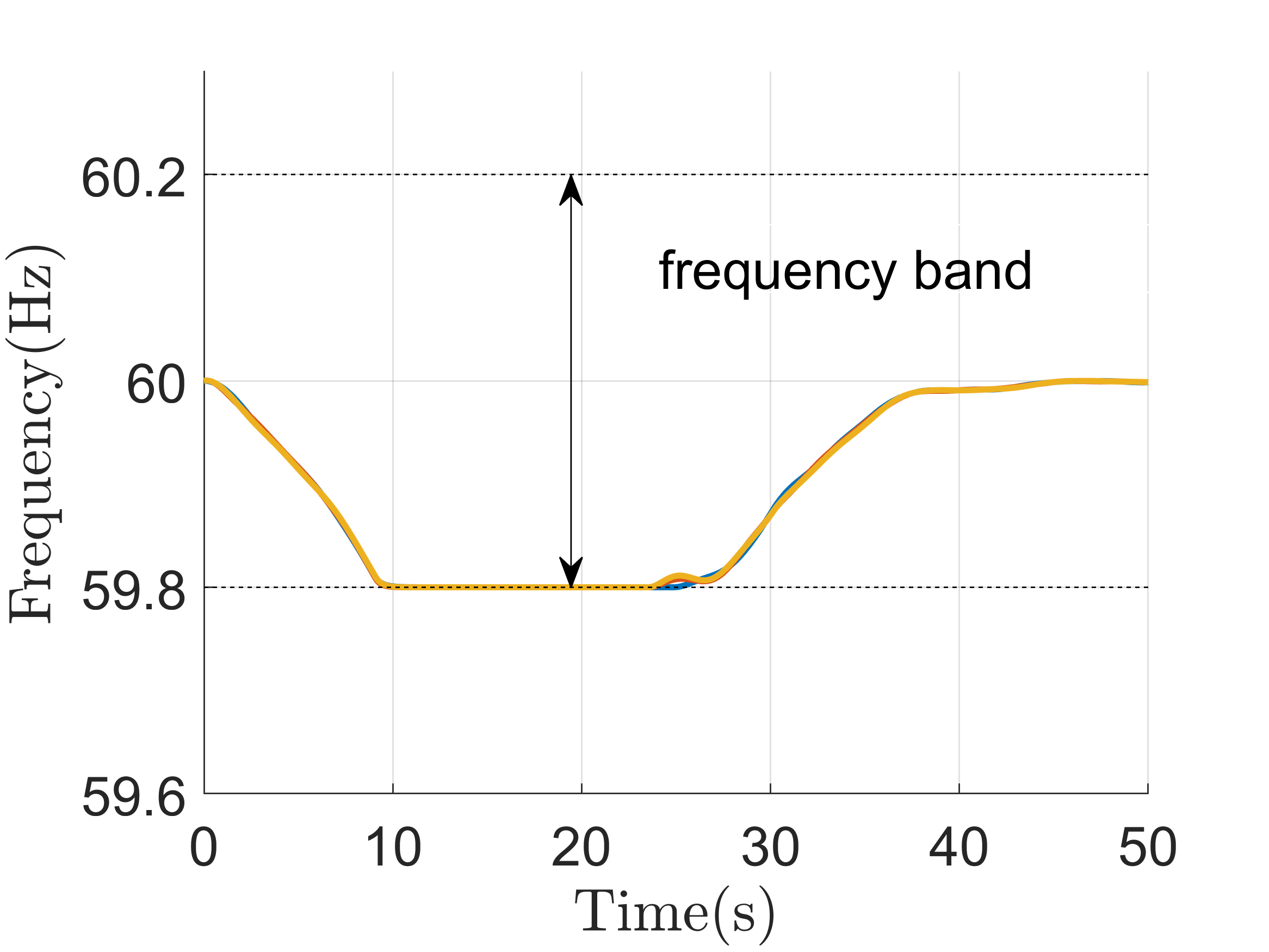}}
  \subfigure[\label{input-trajectories}]{\includegraphics[width=.23\linewidth]{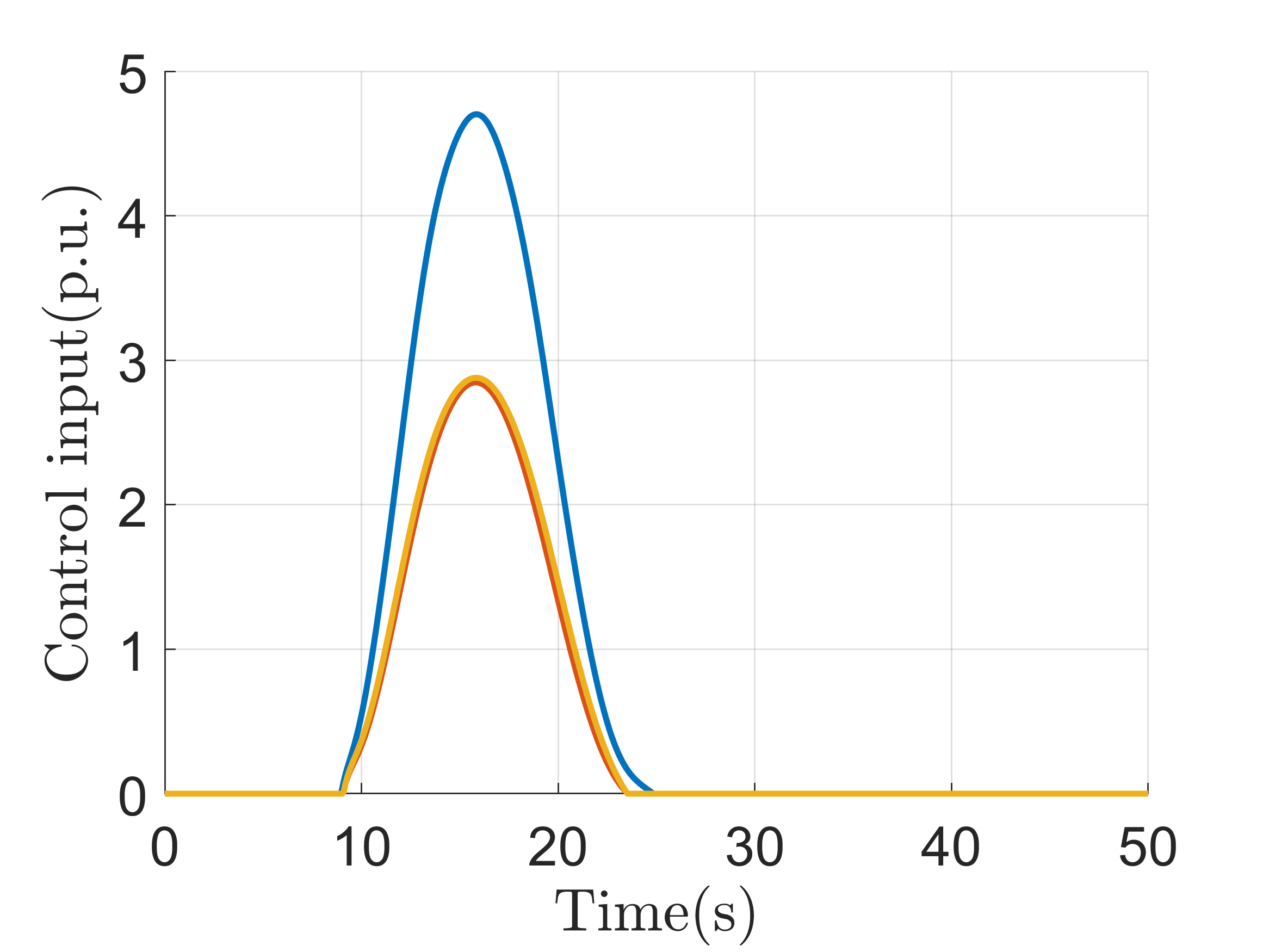}}
  \subfigure[\label{input-trajectories-robust}]{\includegraphics[width=.23\linewidth]{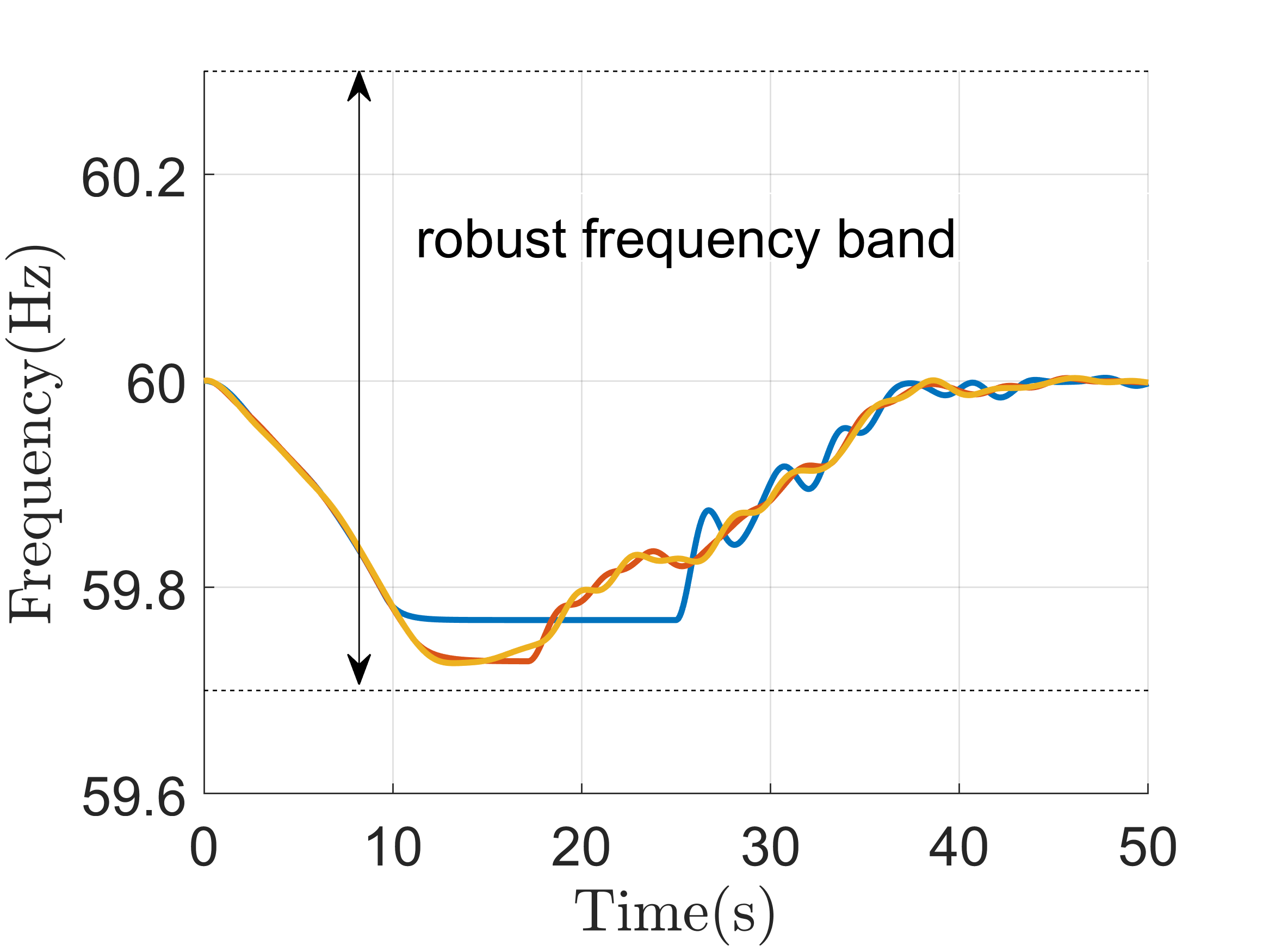}}
  \caption{Frequency and control input trajectories with and without
    transient
    controller. Plot~\subref{frequency-response-no-control-generator}
    shows the frequency trajectories of the generators 30, 31, and 32
    without the transient
    controller~\eqref{eqn:stability-transient-controller-Lipschitz-4},
    with all of them going beyond the lower safe frequency bound. With
    the transient controller,
    plot~\subref{frequency-response-with-control-generator} shows that
    all frequency trajectories stay within the safe
    bound. Plot~\subref{input-trajectories} shows the corresponding
    trajectories of the control
    inputs. Plot~\subref{input-trajectories-robust} shows the
    controller performance under parameter uncertainty and errors in
    the power injection approximation.}\label{fig:trajectories}
\end{figure*}

We first show how the proposed controller maintains the targeted
generator frequencies within the safe region provided that these
frequencies are initially in it. For our first scenario, we consider a
generator loss and recovery process. Specifically, we set the power
injection of node 38 to zero (i.e., generator G9) during the time
interval [10,40]s. As shown in Figure~\ref{fig:generator-loss}, without
the transient
controller~\eqref{eqn:stability-transient-controller-Lipschitz-4}, the
frequency of node 30 first gradually goes down, exceeding the safe
bound 59.8Hz a few times, even tending to converge to a frequency
below it. As node 38 recovers its power supply at 40s, the frequency
comes back to 60Hz. In comparison, with the transient controller, the
frequency trajectory never goes beyond 59.8Hz during the transient.

For our second scenario, we perturb all non-generator nodes by a
sinusoidal power injection whose magnitude is proportional to the
corresponding node's initial power injection. Specifically, for every
$i\in\{1,2,\cdots,29\}$,
\begin{align*}
  p_{i}(t)=
  \begin{cases}
    p_{i}(0) & \text{if $t\geqslant 30$,}
    \\
    \left(1+0.3\sin(\frac{\pi t}{30} )\right)p_{i}(0) &
    \text{otherwise.}
  \end{cases}
\end{align*}
For $i\in\{30,31,\cdots,39\}$, $p_{i}(t)$ remains constant all the
time.
Figure~\ref{fig:trajectories}\subref{frequency-response-no-control-generator}
shows the frequency responses of generators $30$, $31$, and $32$
without the transient controller. One can see that all trajectories
exceed the 59.8Hz lower frequency bound. For comparison,
Figure~\ref{fig:trajectories}\subref{frequency-response-with-control-generator}
shows the trajectories with the transient
controller~(\ref{eqn:stability-transient-controller-Lipschitz-4}),
where all remain within the safe frequency region.
Figure~\ref{fig:trajectories}\subref{input-trajectories} displays the
corresponding input trajectories, which converge to 0 in finite time,
as stated in
Theorem~\ref{thm:decentralized-controller}\emph{\ref{item:finite-time-active}}.
We also illustrate the robustness of the controller against
uncertainty. We have each controller employ $\hat E_{i}=2$ and $\hat
p_{i}(t)=1.1p_{i}(t)$, corresponding to $100\%$ and $10\%$ deviations
on droop coefficients and power injections, respectively.
Figure~\ref{fig:trajectories}\subref{input-trajectories-robust}
illustrates the frequency trajectories of the 3 controlled generators.
Since condition~\eqref{sube:ineq:robust-invariance} is satisfied with
$\Delta=0.1$Hz, Proposition~\ref{prop:robust-uncertainty} ensures that
the invariant frequency interval is now
$[59.7\text{Hz},60.3\text{Hz}]$.

\begin{figure}[tbh!]
  \centering
  \subfigure[\label{IEEE39-omega-trajectories-vs-gamma}]{\includegraphics[width=.47\linewidth]{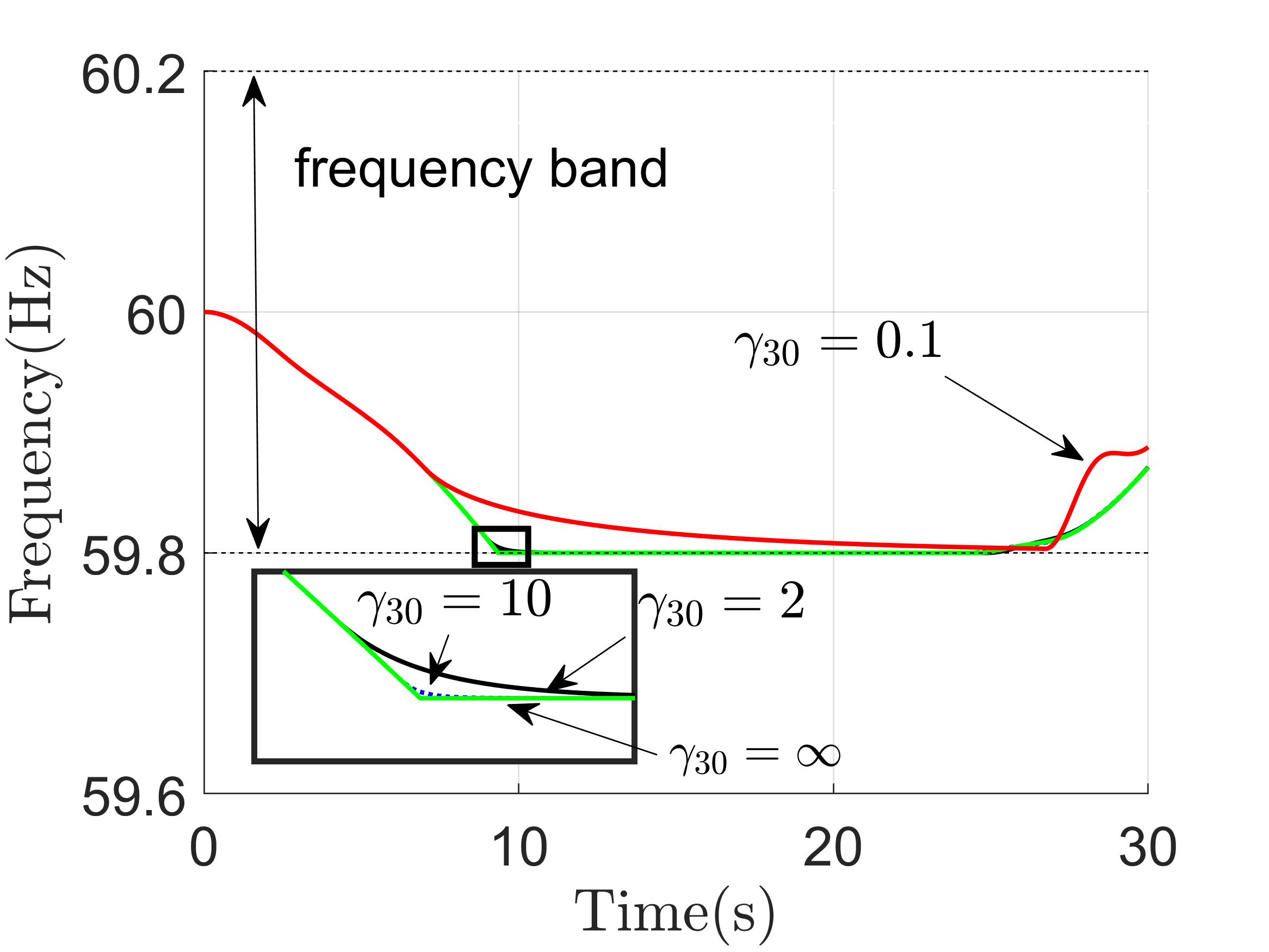}}
  \subfigure[\label{IEEE39-input-trajectories-vs-gamma}]{\includegraphics[width=.47\linewidth]{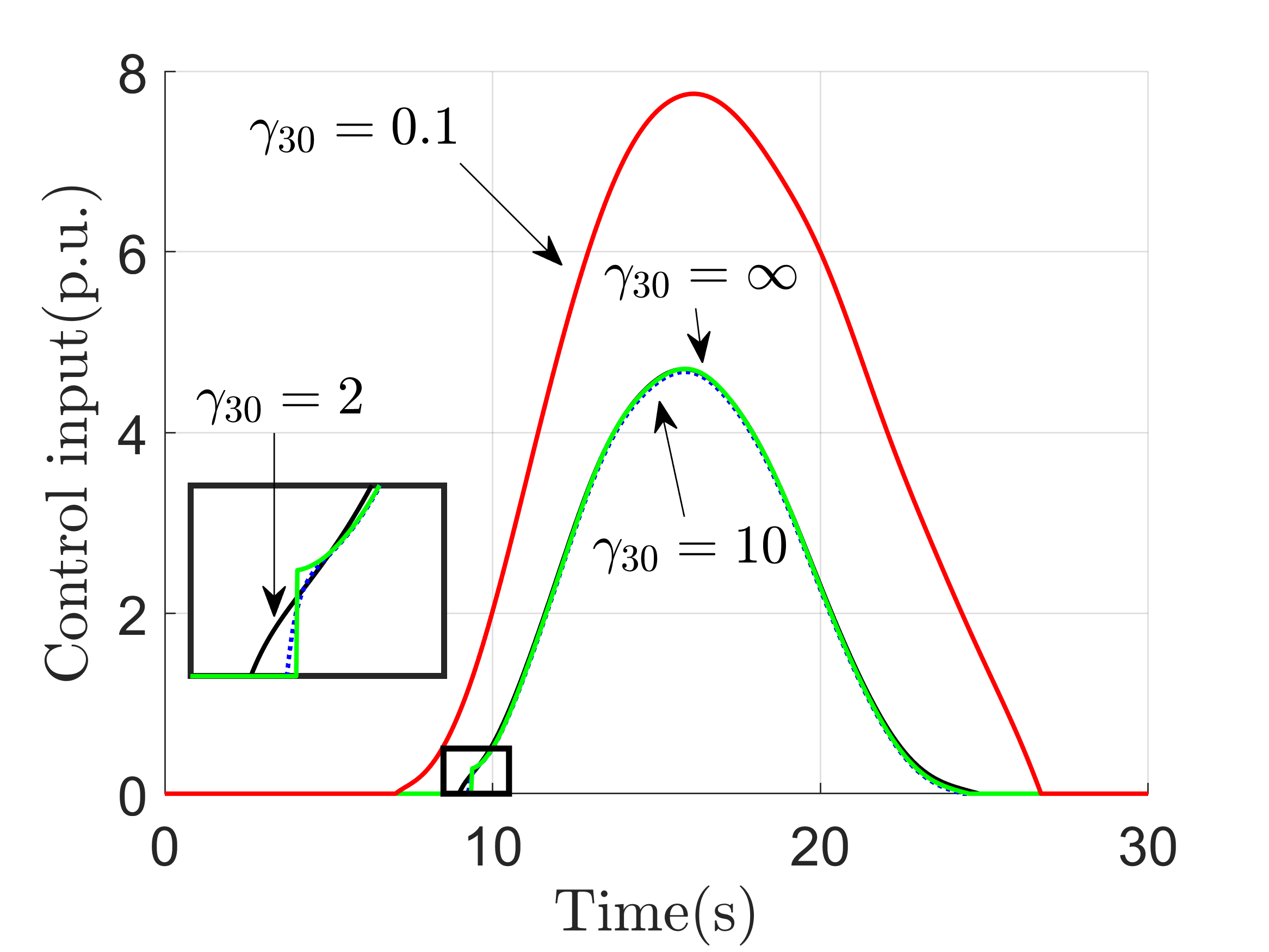}}
  \caption{Frequency and control input trajectories at node $30$ with
    linear class-$\mathcal{K}$ function with slope
    $\gamma_{30}=0.1,2,10$ and $+\infty$, respectively. We observe
    from plot~\subref{IEEE39-omega-trajectories-vs-gamma} that the
    frequency trajectory with small $\gamma_{30}$ tends to stay away
    from the safe frequency bound, at the cost of having a large
    control input, as shown in
    plot~\subref{IEEE39-input-trajectories-vs-gamma}.  A large
    $\gamma_{30}$ causes the controller to be sensitive to
    $\omega_{30}$, making the input change rapidly around 9s.
  }\label{fig:trajectories-vs-gamma}
\end{figure}

\begin{figure}[tbh!]
  \centering
  \subfigure[\label{IEEE39-input-noisy-trajectories-gamma-2}]{\includegraphics[width=.47\linewidth]{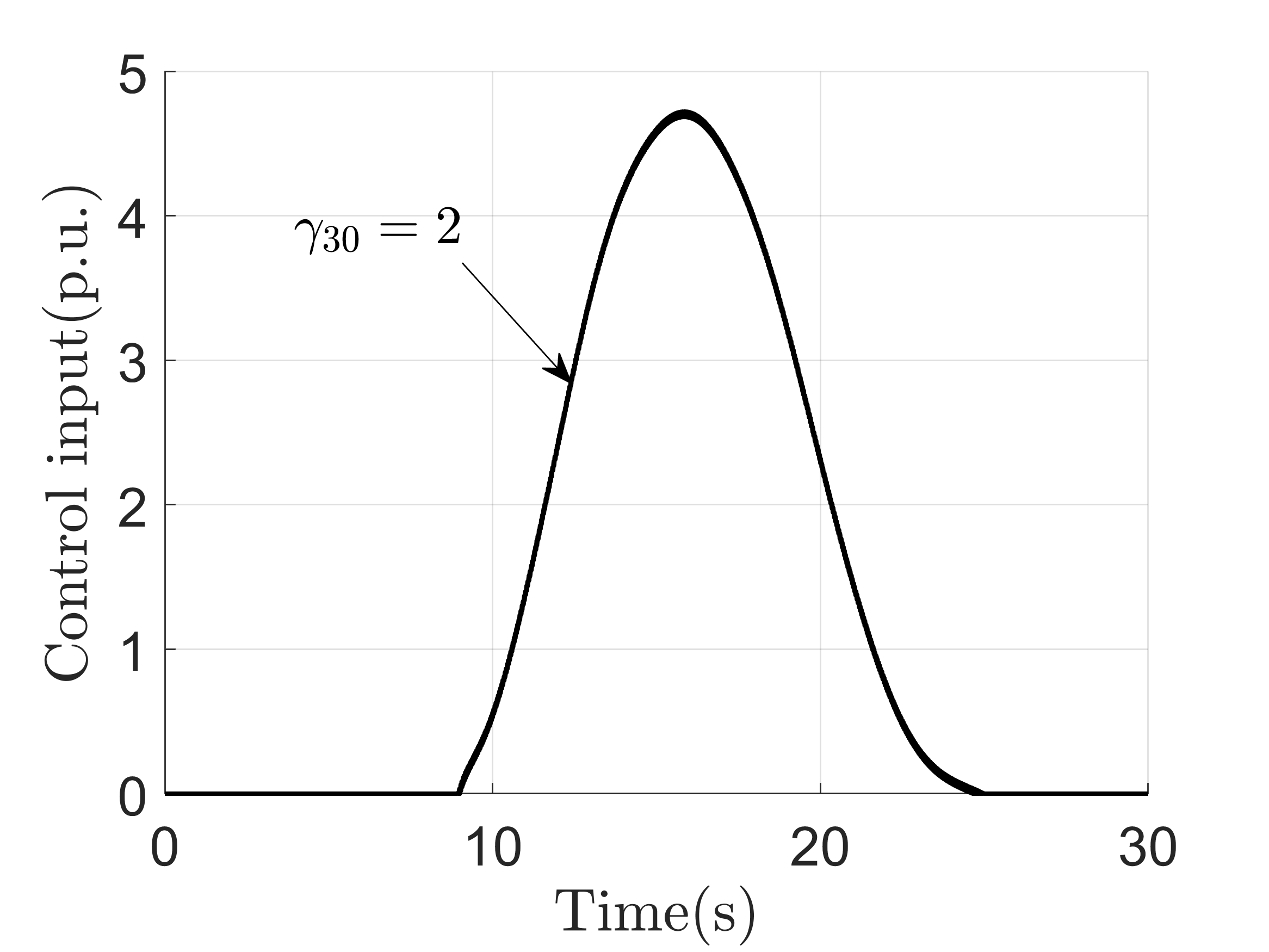}}
  \subfigure[\label{IEEE39-input-noisy-trajectories-gamma-infty}]{\includegraphics[width=.47\linewidth]{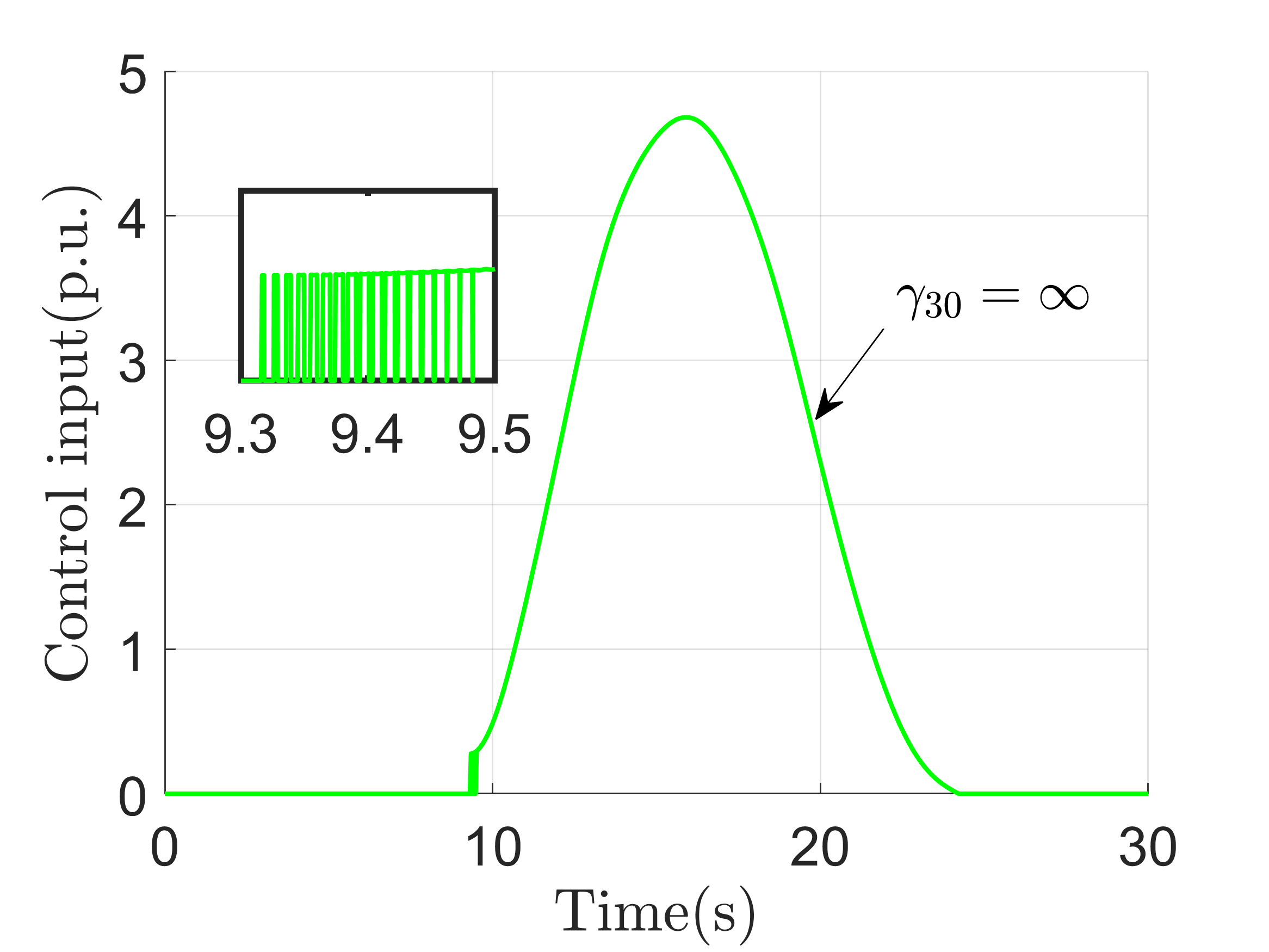}}
  \caption{Control input trajectories at node $30$ with linear
    class-$\mathcal{K}$ function with slope $\gamma_{30}=2$ and
    $+\infty$, respectively, under state measurement errors in
    $\omega_{30}$.  The controller with $\gamma_{30}=2$ is Lipschitz
    continuous (cf.
    plot~\subref{IEEE39-input-noisy-trajectories-gamma-2}), whereas
    the controller with $\gamma_{30}=+\infty$ (cf.
    plot~\subref{IEEE39-input-noisy-trajectories-gamma-infty}) is
    discontinuous.}\label{fig:trajectories-vs-noisy-gamma}
\end{figure}

Next, we examine the effect of the choice of class-$\mathcal{K}$
function on the behavior of the transient frequency.  We focus our
attention on bus $30$ and simulate the network behavior for a linear
function with $\gamma_{30}=0.1,2,10,$ and $+\infty$ (the latter
corresponding to the discontinuous controller
in~\eqref{eqn:stability-transient-controller-infinite}).
Figure~\ref{fig:trajectories-vs-gamma} shows the corresponding
frequency and control input trajectories for the first 30 seconds at
node 30. From
Figure~\ref{fig:trajectories-vs-gamma}\subref{IEEE39-omega-trajectories-vs-gamma},
one can see that the frequency trajectory with $\gamma_{30}=0.1$ tends
to stay away from the lower safe bound (overprotection), compared with
the trajectories with $\gamma_{30}=2,10,$ and $+\infty$, and this
results in a larger control input, cf.
Figure~\ref{fig:trajectories-vs-gamma}\subref{IEEE39-input-trajectories-vs-gamma}.
As $\gamma_{30}$ increases, the control input is triggered later.  On
the other hand, choosing a large $\gamma_{30}$ lead to higher
sensitivity, as observed in
Figure~\ref{fig:trajectories-vs-gamma}\subref{IEEE39-input-trajectories-vs-gamma},
where the input trajectory with large $\gamma_{30}$ grows faster at
the time when the control input first becomes non-zero. In fact, the
controller with $\gamma_{30}=10$ exhibits a sharp change around
$t=9s$, similar to the discontinuous
controller~\eqref{eqn:stability-transient-controller-infinite}. The
discontinuity of the latter is more evident under state measurements
errors.  In Figure~\ref{fig:trajectories-vs-noisy-gamma}, we run the
same simulation but with
$\hat\omega_{30}(t)=\omega_{30}(t)+0.001\sin(200\pi t)$ as the
measured frequency. One can observe the high-frequency fluctuation in
the control input trajectory around $9.4$s for $\gamma_{30}=+\infty$,
whereas this does not happen for $\gamma_{30}=2$ due to its Lipschitz
continuity character.  These simulations validate the observations of
Remark~\ref{rmk:linear-class-K}.

\begin{figure}[tbh!]
  \centering
  \subfigure[\label{freuency-response-with-delayed-control-generator}]{\includegraphics[width=.47\linewidth]{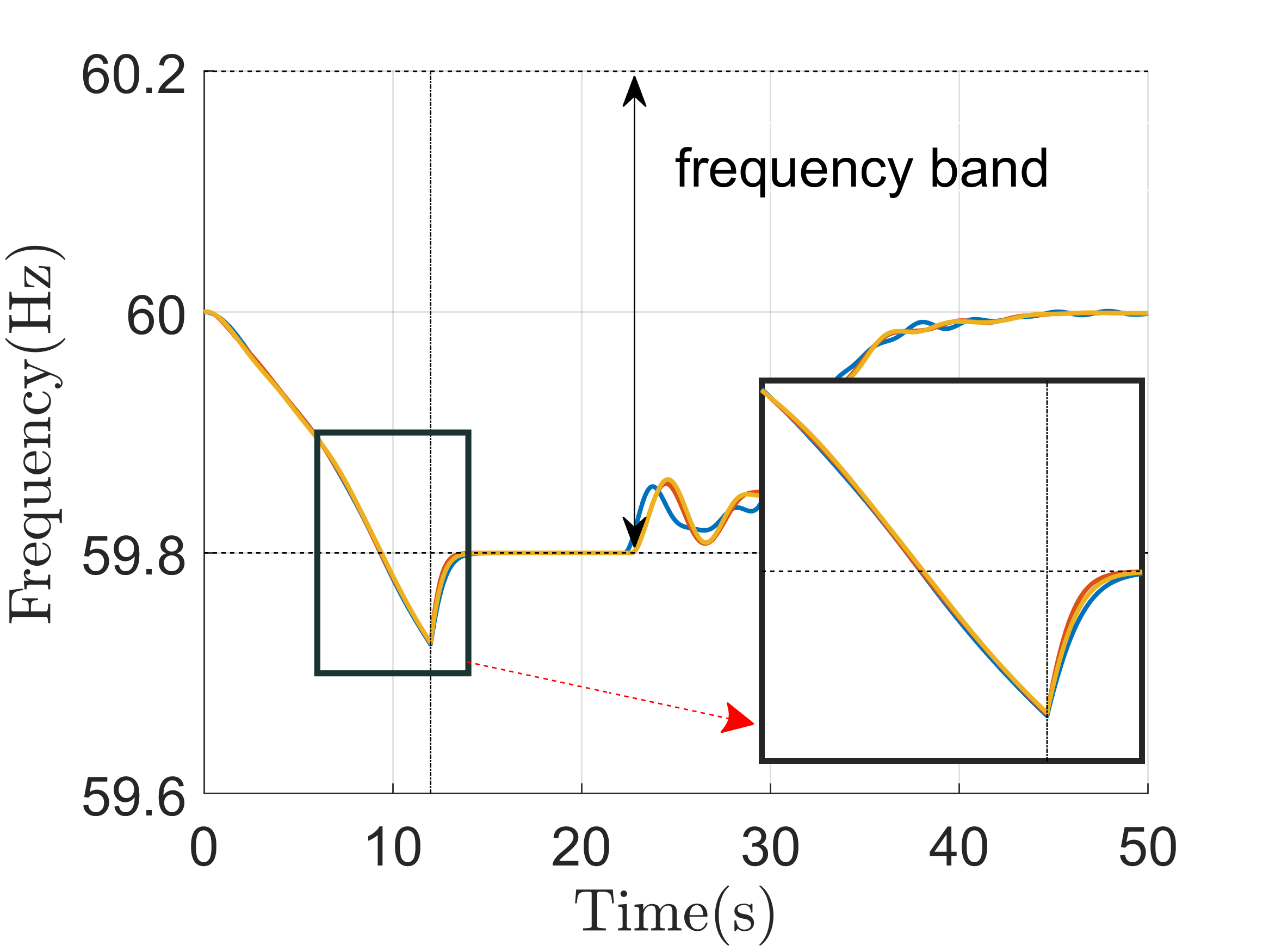}} 
  \subfigure[\label{delayed-input-trajectories}]{\includegraphics[width=.47\linewidth]{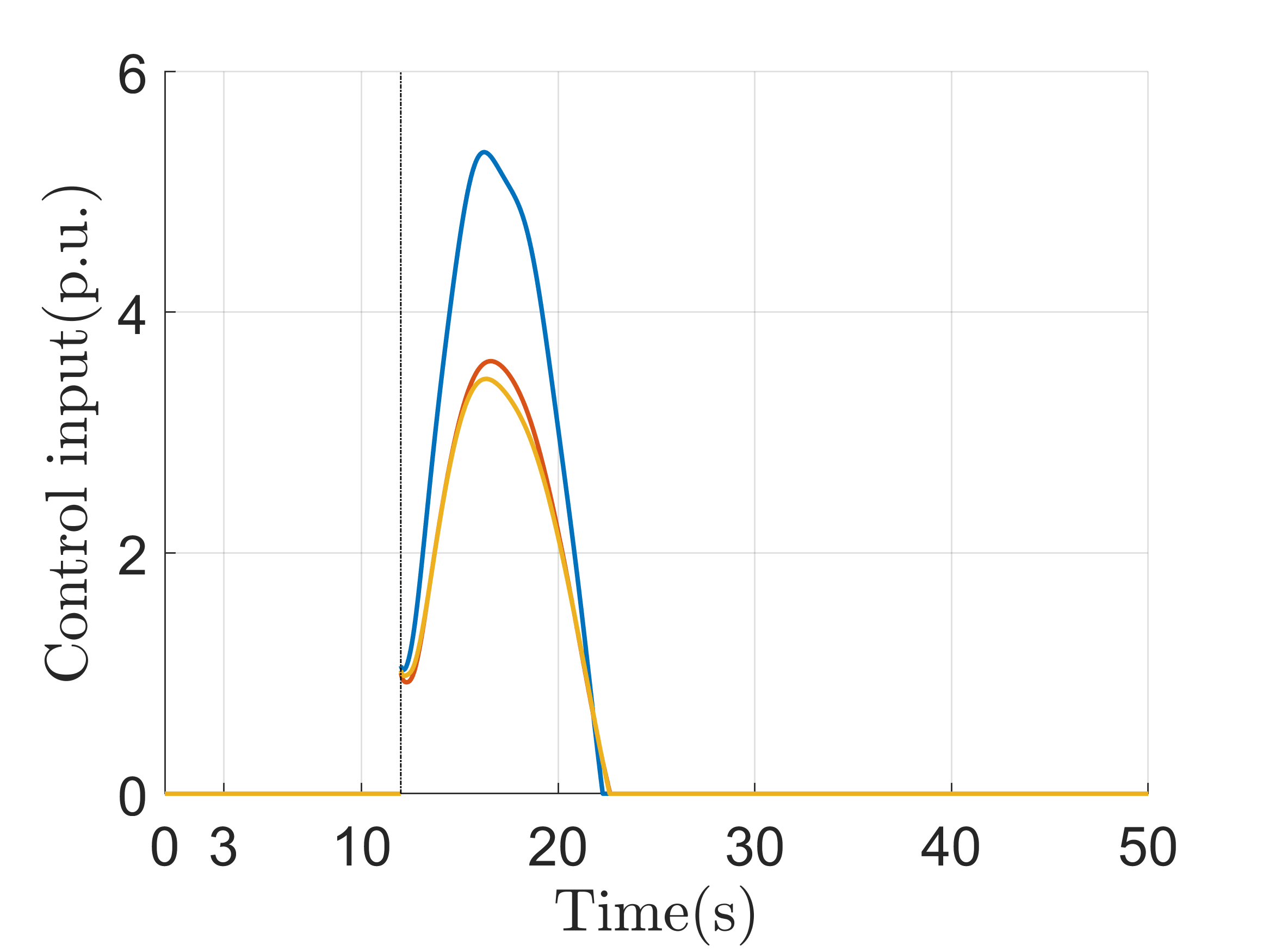}}
  \caption{Frequency and control input trajectories with transient
    controller available after
    $t=12$s. Plot~\subref{freuency-response-with-delayed-control-generator}
    shows the frequency trajectories of generators 30, 31, and 32. Due
    to the disturbance, and without the transient controller, all 3
    frequency trajectories exceed the 59.8Hz safe bound at $t=12$s.
    As the transient controller kicks in, the unsafe trajectories come
    back to the safe region and never leave
    afterwards. Plot~\subref{delayed-input-trajectories} shows the
    control input trajectories.}\label{fig:delayed-trajectories}
\end{figure}

Next, we simulate the case where some of the generator frequencies are
initially outside the safe region to show how the transient controller
brings the frequencies back to it.  We use the same setup as in
Figure~\ref{fig:trajectories}, but we only turn on the distributed
controller
after~$t=12s$. Figure~\ref{fig:delayed-trajectories}\subref{freuency-response-with-delayed-control-generator}
shows the frequency trajectories of generators $30$, $31$, and
$32$. As the controller is disabled for the first $12$s, all 3
frequency trajectories are lower than 59.8hz at $t=12$s.  After
$t=12$s, all of them return to the safe region in a monotonic way, and
once they are in the region, they never leave, in accordance with
Theorem~\ref{thm:decentralized-controller}\emph{\ref{item:frequency-attraction}}.
Figure~\ref{fig:delayed-trajectories}\subref{delayed-input-trajectories}
shows the corresponding control input trajectories.

\begin{figure}[htb]
  \centering%
  \subfigure[\label{control-bound-lower}]{\includegraphics[width=.47\linewidth]{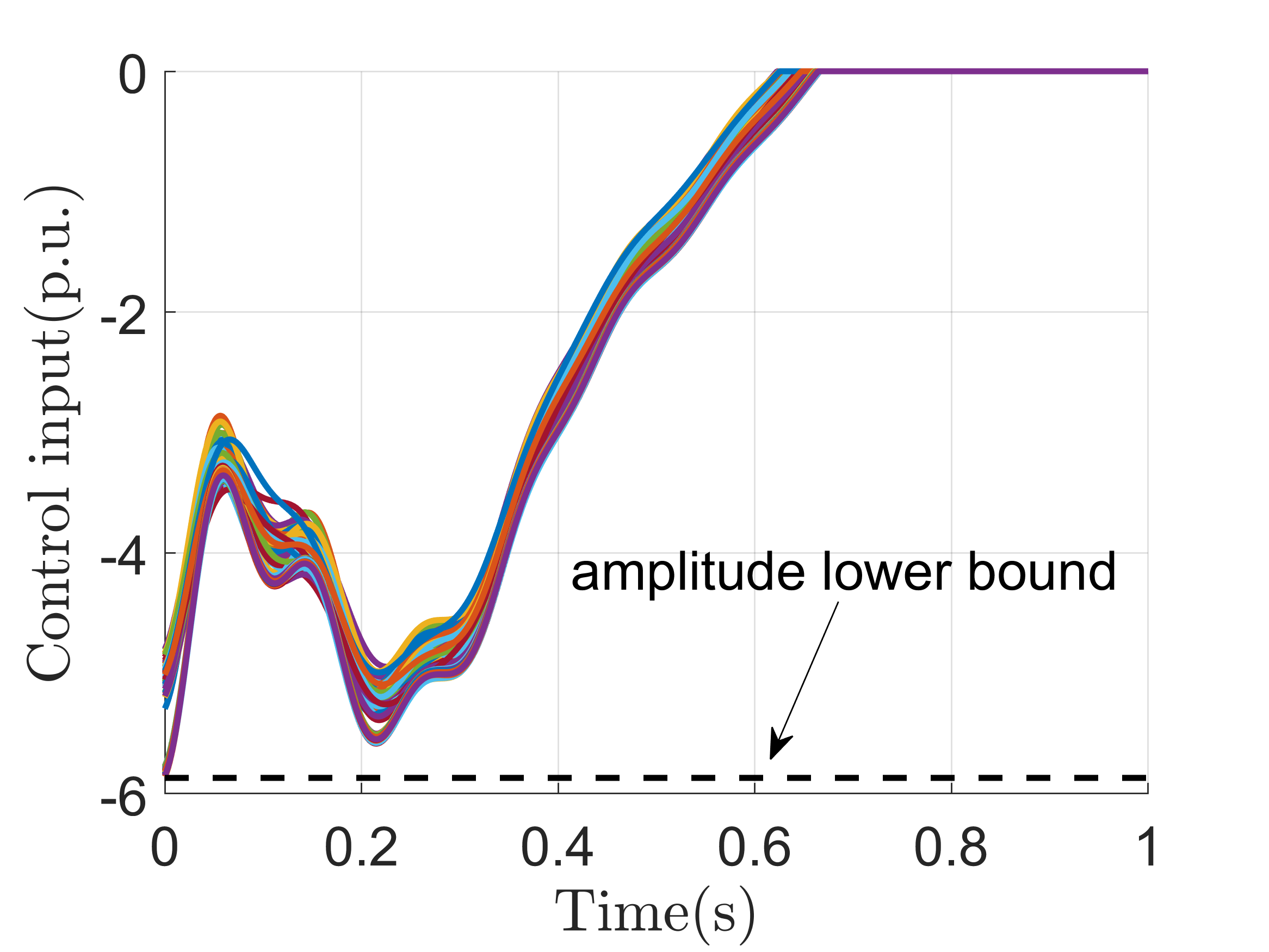}}
  \subfigure[\label{control-bound-upper}]{\includegraphics[width=.47\linewidth]{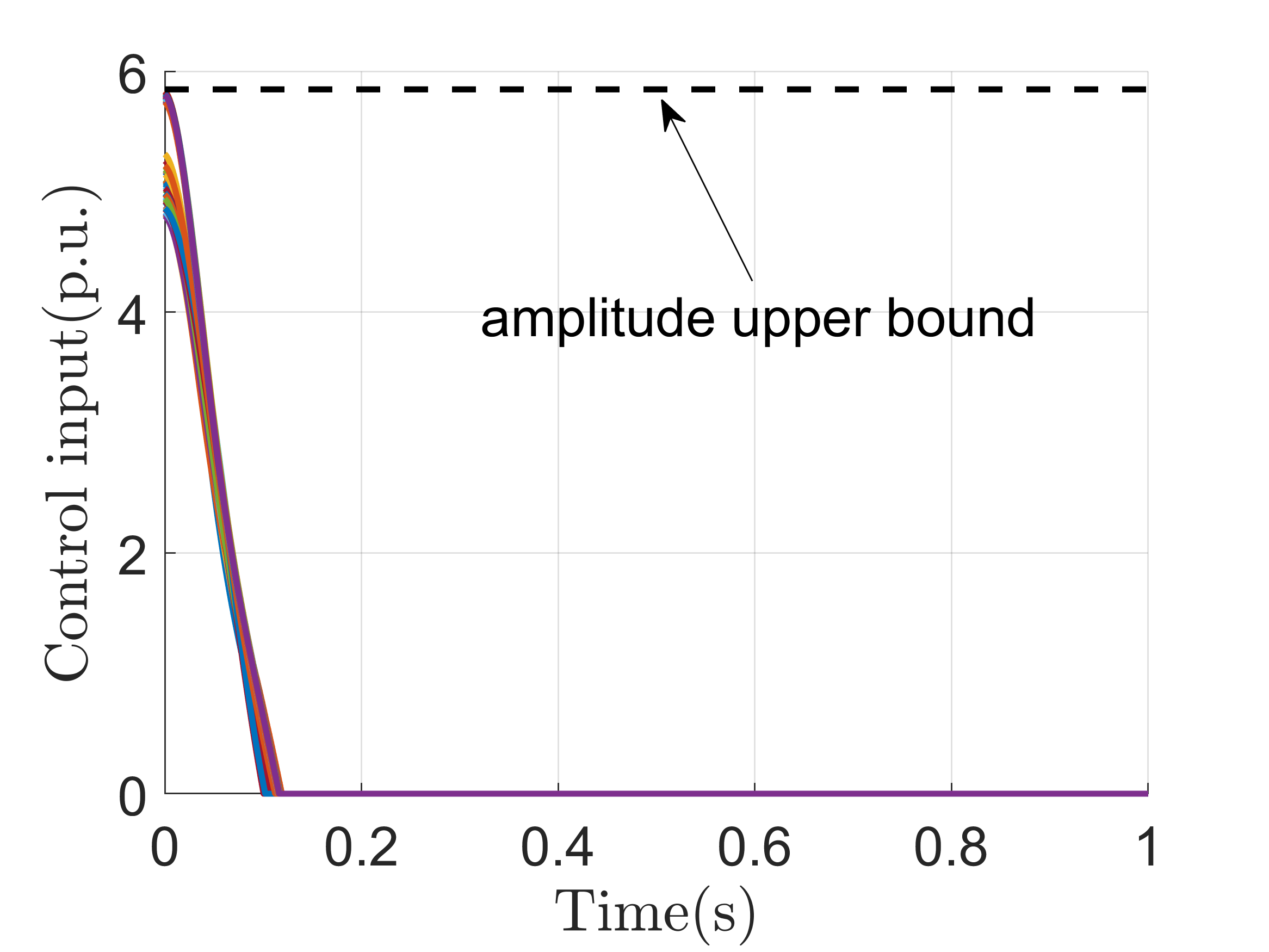}}
  \caption{Control input trajectories at node 30 corresponding to 100
    different initial states. In plot~\subref{control-bound-lower},
    with all initial states randomly selected around the worst-case
    scenario, all 100 trajectories are lower bounded by $-5.8686$
    (denoted by the dashed line), as guaranteed by
    Lemma~\ref{lemma:lower-bound}.  A similar result is illustrated in
    plot~\subref{control-bound-upper}, where another 100 trajectories
    with random initial states are upper bounded by
    $5.8494$.}\label{fig:IEEE39bus-control-bnd}
\end{figure}

Finally, we illustrate the bounds on control amplitude of
Section~\ref{subsection:magnitude}. Let $\eta=0.5$ and $i=30$. By
Lemma~\ref{lemma:lower-bound}, the control input is lower bounded by
$u_{i}^{\min}(\gamma)$, which requires
$g_{i}(\lambda^{*},\omega^{*})$.  The numerical computation of the
upper $g_{i}(\lambda^{o},\omega^{o})$ (cf.
Lemma~\ref{lemma:upper-optimal}) and lower $h_{i}(\underline
z^{\mu^{*}} ,\underline\omega^{\mu^{*}})$ (cf.
Lemma~\ref{lemma:lower-optimal}) bounds both yield $-5.8686$.
Figure.~\ref{fig:IEEE39bus-control-bnd}\subref{control-bound-lower}
shows 100 input trajectories with initial states randomly selected
around $(\lambda^{o},\omega^{o})$, all lower bounded by $-5.8686$.

\section{Conclusions}
We have proposed a distributed transient power frequency controller
that is able to maintain the nodal frequency of actuated buses within
a desired safe region and to recover from undesired initial
conditions. We have proven that the control input vanishes in finite
time, so that the closed-loop system possesses the same equilibrium
and local stability and convergence guarantees as the open-loop one.
We have characterized the smoothness and robustness properties of the
proposed controller.  Future work will investigate the incorporation
of economic cost, taking advantage of the trade-offs in the choice of
class-$\mathcal{K}$ functions for controller design, the optimization
of control effort by having controlled nodes have access to
information beyond their immediate neighbors, and the understanding of
the connection between actuation effort and network connectivity.

{\small
\bibliographystyle{plainnat}
\bibliography{alias,JC,Main,Main-add}

\begin{thebibliography}{26}
\providecommand{\natexlab}[1]{#1}
\providecommand{\url}[1]{\texttt{#1}}
\expandafter\ifx\csname urlstyle\endcsname\relax
  \providecommand{\doi}[1]{doi: #1}\else
  \providecommand{\doi}{doi: \begingroup \urlstyle{rm}\Url}\fi

\bibitem[Alam and Makram(2006)]{AA-EBM:06}
A.~Alam and E.B. Makram.
\newblock Transient stability constrained optimal power flow.
\newblock In \emph{IEEE Power and Energy Society General Meeting}, Montreal,
  Canada, June 2006.
\newblock Electronic proceedings.

\bibitem[Althoff(2014)]{MA:14}
M.~Althoff.
\newblock Formal and compositional analysis of power systems using reachable
  sets.
\newblock \emph{IEEE Transactions on Power Systems}, 29\penalty0 (5):\penalty0
  2270--2280, 2014.

\bibitem[Ames et~al.(2017)Ames, Xu, Grizzle, and Tabuada]{ADA-XX-JWG-PT:16}
A.~D. Ames, X.~Xu, J.~W. Grizzle, and P.~Tabuada.
\newblock Control barrier function based quadratic programs for safety critical
  systems.
\newblock \emph{IEEE Transactions on Automatic Control}, 62\penalty0
  (8):\penalty0 3861--3876, 2017.

\bibitem[Biggs(1994)]{NB:94}
N.~Biggs.
\newblock \emph{Algebraic Graph Theory}.
\newblock Cambridge University Press, 2 edition, 1994.
\newblock ISBN 0521458978.

\bibitem[Blancini and Miani(2008)]{FB-SM:08}
F.~Blancini and S.~Miani.
\newblock \emph{Set-theoretic Methods in Control}.
\newblock Birkh{\"a}user, Boston, MA, 2008.

\bibitem[Borsche et~al.(2015)Borsche, Liu, and Hill]{TSB-TL-DJH:15}
T.~S. Borsche, T.~Liu, and D.~J. Hill.
\newblock Effects of rotational inertia on power system damping and frequency
  transients.
\newblock In \emph{{IEEE} Conf.\ on Decision and Control}, pages 5940--5946,
  Osaka, Japan, 2015.

\bibitem[Bullo et~al.(2009)Bullo, Cort{\'e}s, and
  Mart{\'\i}nez]{FB-JC-SM:08cor}
F.~Bullo, J.~Cort{\'e}s, and S.~Mart{\'\i}nez.
\newblock \emph{Distributed Control of Robotic Networks}.
\newblock Applied Mathematics Series. Princeton University Press, 2009.
\newblock ISBN 978-0-691-14195-4.
\newblock Electronically available at \url{http://coordinationbook.info}.

\bibitem[Chen and Dom{\'i}nguez-Garc{\'i}a(2012)]{YCC-ADD:12}
Y.~C. Chen and A.~D. Dom{\'i}nguez-Garc{\'i}a.
\newblock A method to study the effect of renewable resource variability on
  power system dynamics.
\newblock \emph{IEEE Transactions on Power Systems}, 27\penalty0 (4):\penalty0
  1978--1989, 2012.

\bibitem[Cheung et~al.(2009)Cheung, Chow, and Rogers]{KWC-JC-GR:09}
K.~W. Cheung, J.~Chow, and G.~Rogers.
\newblock \emph{Power System Toolbox, v 3.0.}
\newblock Rensselaer Polytechnic Institute and Cherry Tree Scientific Software,
  2009.

\bibitem[Chiang(2011)]{HDC:11}
H.~D. Chiang.
\newblock \emph{Direct Methods for Stability Analysis of Electric Power
  Systems: Theoretical Foundation, BCU Methodologies, and Applications}.
\newblock John Wiley and Sons, 2011.

\bibitem[Choi et~al.(2016)Choi, Seiler, and Dhople]{HC-PJS-SVD:16}
H.~Choi, P.~J. Seiler, and S.~V. Dhople.
\newblock Propagating uncertainty in power-system {DAE} models with
  semidefinite programming.
\newblock \emph{IEEE Transactions on Power Systems}, 32\penalty0 (4):\penalty0
  3146--3156, 2016.

\bibitem[D{\"o}rfler et~al.(2013)D{\"o}rfler, Chertkov, and Bullo]{FD-MC-FB:13}
F.~D{\"o}rfler, M.~Chertkov, and F.~Bullo.
\newblock Synchronization in complex oscillator networks and smart grids.
\newblock \emph{Proceedings of the National Academy of Sciences}, 110\penalty0
  (6):\penalty0 2005--2010, 2013.

\bibitem[Grunbaum and Pernot(2001)]{TG-JP:01}
R.~Grunbaum and J.~Pernot.
\newblock Thyristor-controlled series compensation: A state of the art approach
  for optimization of transmission over power links.
\newblock Technical report, ABB, 2001.

\bibitem[Khalil(2002)]{HKK:02}
H.~K. Khalil.
\newblock \emph{Nonlinear Systems}.
\newblock Prentice Hall, 3 edition, 2002.
\newblock ISBN 0130673897.

\bibitem[Kundur(1994)]{PK:94}
P.~Kundur.
\newblock \emph{Power System Stability and Control}.
\newblock McGraw-Hill, 1994.
\newblock ISBN 007035958X.

\bibitem[Kundur et~al.(2004)Kundur, Paserba, Ajjarapu, Andersson, Bose,
  Canizares, Hatziargyriou, Hill, Stankovic, Taylor, Cutsem, and
  Vittal]{PK-JP:04}
P.~Kundur, J.~Paserba, V.~Ajjarapu, G.~Andersson, A.~Bose, C.~Canizares,
  N.~Hatziargyriou, D.~Hill, A.~Stankovic, C.~Taylor, T.~V. Cutsem, and
  V.~Vittal.
\newblock Definition and classification of power system stability.
\newblock \emph{IEEE Transactions on Power Systems}, 19\penalty0 (2):\penalty0
  1387--1401, 2004.

\bibitem[Mahmud et~al.(2014)Mahmud, Pota, Aldeen, and
  Hossain]{MAM-HRP-MA-MJH:14}
M.~A. Mahmud, H.~R. Pota, M.~Aldeen, and M.~J. Hossain.
\newblock Partial feedback linearizing excitation controller for multimachine
  power systems to improve transient stability.
\newblock \emph{IEEE Transactions on Power Systems}, 29:\penalty0 561--571,
  2014.

\bibitem[Menck et~al.(2014)Menck, Heitzig, Kurths, and
  Schellnhuber]{PJM-JH-JK-HJS:14}
P.~J. Menck, J.~Heitzig, J.~Kurths, and H.~J. Schellnhuber.
\newblock How dead ends undermine power grid stability.
\newblock \emph{Nature Communications}, 5\penalty0 (3969):\penalty0 1--8, 2014.

\bibitem[Miller et~al.(2011)Miller, Clark, and Shao]{NWM-KC-MS:11}
N.~W. Miller, K.~Clark, and M.~Shao.
\newblock Frequency responsive wind plant controls: Impacts on grid
  performance.
\newblock In \emph{Power and Energy Society General Meeting ({PESGM})}, pages
  1--8, 2011.

\bibitem[Nguyen et~al.(2011)Nguyen, Nguyen, and Karimishad]{TTN-VLN-AK:11}
T.~T. Nguyen, V.~L. Nguyen, and A.~Karimishad.
\newblock Transient stability-constrained optimal power flow for online
  dispatch and nodal price evaluation in power systems with flexible ac
  transmission system devices.
\newblock \emph{IET Generation, Transmission \& Distribution}, 5:\penalty0
  332--346, 2011.

\bibitem[Poolla et~al.(2017)Poolla, Bolognani, and Dorfler]{BKP-SB-FD:17}
B.~K. Poolla, S.~Bolognani, and F.~Dorfler.
\newblock Optimal placement of virtual inertia in power grids.
\newblock \emph{IEEE Transactions on Automatic Control}, 2017.
\newblock \doi{10.1109/TAC.2017.2703302}.
\newblock To appear.

\bibitem[Pouyan et~al.(2006)Pouyan, Kundur, and Taylor]{PP-PSK-CWT:06}
P.~Pouyan, P.~S. Kundur, and C.~W. Taylor.
\newblock The anatomy of a power grid blackout-root causes and dynamics of
  recent major blackouts.
\newblock \emph{IEEE Power and Energy Magazine}, 4\penalty0 (5):\penalty0
  22--29, 2006.

\bibitem[Vu et~al.(2017)Vu, Nguyen, Megretski, Slotine, and
  Turitsyn]{TLV-HDN-AM-JS-KT:17}
T.~L. Vu, H.~D. Nguyen, A.~Megretski, J.~Slotine, and K.~Turitsyn.
\newblock Inverse stability problem and applications to renewables integration.
\newblock 2017.
\newblock https://arxiv.org/pdf/1703.04491.pdf.

\bibitem[Vu et~al.(2018)Vu, Nguyen, Megretski, Slotine, and
  Turitsyn]{TLV-HDN-AM-JS-KT:18}
T.~L. Vu, H.~D. Nguyen, A.~Megretski, J.~Slotine, and K.~Turitsyn.
\newblock Inverse stability problem and applications to renewables integration.
\newblock \emph{IEEE Control Systems Letters}, 2\penalty0 (1):\penalty0
  133--138, 2018.

\bibitem[Zhang and Cort\'es(2017)]{YZ-JC:17-acc}
Y.~Zhang and J.~Cort\'es.
\newblock Transient-state feasibility set approximation of power networks
  against disturbances of unknown amplitude.
\newblock In \emph{{A}merican {C}ontrol {C}onference}, pages 2767--2772,
  Seattle, WA, May 2017.

\bibitem[Zhang and Cort\'{e}s(2018)]{YZ-JC:18-cdc1}
Y.~Zhang and J.~Cort\'{e}s.
\newblock Distributed transient frequency control in power networks.
\newblock In \emph{{IEEE} Conf.\ on Decision and Control}, Miami Beach, FL,
  December 2018.
\newblock To appear.

\end{thebibliography}
}

\end{document}